\documentclass[twocolumn]{aastex631}
\usepackage{amsmath}

\shorttitle{Quasar UV/optical damping timescale}
\shortauthors{Zhou et al.}

\graphicspath{{./}{figures/}}

\begin{document}

\title{How long will the quasar UV/optical flickering be damped?}

\correspondingauthor{Mouyuan Sun}
\email{msun88@xmu.edu.cn}

\author[0009-0005-2801-6594]{Shuying Zhou}
\affiliation{Department of Astronomy, Xiamen University, Xiamen, Fujian 361005, China; msun88@xmu.edu.cn}

\author[0000-0002-0771-2153]{Mouyuan Sun}
\affiliation{Department of Astronomy, Xiamen University, Xiamen, Fujian 361005, China; msun88@xmu.edu.cn}

\author[0000-0002-4223-2198]{Zhen-Yi Cai}
\affiliation{Department of Astronomy, University of Science and Technology of China, Hefei 230026, China}
\affiliation{School of Astronomy and Space Science, University of Science and Technology of China, Hefei 230026, China}

\author[0000-0002-1497-8371]{Guowei Ren}
\affiliation{Department of Astronomy, Xiamen University, Xiamen, Fujian 361005, China; msun88@xmu.edu.cn}

\author[0000-0002-4419-6434]{Jun-Xian Wang}
\affiliation{Department of Astronomy, University of Science and Technology of China, Hefei 230026, China}
\affiliation{School of Astronomy and Space Science, University of Science and Technology of China, Hefei 230026, China}

\author[0000-0002-1935-8104]{Yongquan Xue}
\affiliation{Department of Astronomy, University of Science and Technology of China, Hefei 230026, China}
\affiliation{School of Astronomy and Space Science, University of Science and Technology of China, Hefei 230026, China}

\begin{abstract}

The UV/optical light curves of Active Galactic Nuclei (AGNs) are commonly described by the Damped Random Walk (DRW) model. However, the physical interpretation of the damping timescale, a key parameter in the DRW model, remains unclear. Particularly, recent observations indicate a weak dependence of the damping timescale upon both wavelength and accretion rate, clearly being inconsistent with the accretion-disk theory. In this study, we investigate the damping timescale in the framework of the Corona Heated Accretion disk Reprocessing (CHAR) model, a physical model that describes AGN variability. We find that while the CHAR model can reproduce the observed power spectral densities of the 20-year light curves for 190 sources from \cite{Stone2022}, the observed damping timescale, as well as its weak dependence on wavelength, can also be well recovered through fitting the mock light curves with DRW. We further demonstrate that such weak dependence is artificial due to the effect of inadequate durations of light curves, which leads to best-fitting damping timescales lower than the intrinsic ones. After eliminating this effect, the CHAR model indeed yields a strong dependence of the intrinsic damping timescale on the bolometric luminosity and rest-frame wavelength. Our results highlight the demand for sufficiently long light curves in AGN variability studies and important applications of the CHAR model in such studies. 

\end{abstract}


\section{Introduction} \label{sec: intro}

At the centers of Active Galactic Nuclei (AGNs), growing supermassive black holes (SMBHs) are surrounded by gaseous structures such as accretion disks, broad-line regions, and dust tori. It is widely believed that most of the intrinsic properties of AGNs stem from the accretion effect of SMBHs, such as high luminosities and intense electromagnetic radiations in the entire electromagnetic spectrum. The emission of AGNs exhibits a complex and significant stochastic variability over a wide range of wavelengths, e.g., in the ultraviolet (UV) and optical bands \citep[e.g.,][]{Ulrich1997}. This inherent variability provides a unique way to understand AGNs' structures and intrinsic physical mechanisms. It has been shown that there are relationships between the AGN UV/optical fractional variability amplitude and the AGN luminosity \citep[e.g.,][]{MacLeod2010, Zuo2012, Morganson2014, Li2018-DES, Sun2018, Suberlak2021}, Eddington ratio \citep[e.g.,][]{MacLeod2010, Morganson2014, Simm2016, Sun2018, DeCicco2022}, rest-frame wavelength \citep[e.g.,][]{MacLeod2010, Morganson2014, Sun2015, Simm2016, Sanchez2018, Suberlak2021}, and other relevant parameters \citep[e.g.,][]{Ai2010, MacLeod2010, Sun2018, Kang2018}. By measuring the time delay between the variability of different bands using the reverberation mapping technique \citep[][]{Blandford1982}{}{}, one can measure the size of the AGN accretion disk and the broad-line region, helping us estimate the virial mass of the central SMBH and test the accretion-disk theory \citep[e.g.,][]{Fausnaugh2016, Du2019, Cackett2021}. 

Although much progress has been achieved in the studies of AGN variability, the primary physical mechanism causing variability is still an unresolved issue. Many theories have been proposed, such as the global variation of the accretion rate \citep{Lyubarskii1997, Li2008, Liu2016}, the local temperature fluctuation \citep{Kelly2009, Dexter2011, Cai2016}, and the effect of large-scale fluctuations on local temperature fluctuations \citep{Cai2018, Neustadt2022, Secunda2023}. However, these theories failed to fully account for observational results, e.g., AGN power spectral densities (PSDs). More sophisticated theoretical models and better observational data are required to investigate AGN variability further. 

The Damped Random Walk (DRW) model is an effective statistical model for fitting AGN light curves \citep[e.g.,][]{Kelly2009, Kozlowski2010, MacLeod2010, Zu2013, Suberlak2021}. Its power spectral density is described by a power-law function $f^{-2}$ at high frequencies, which transits to white noise at low frequencies. The transition frequency is $f_0=1/(2\pi \tau_{\mathrm{DRW}})$, where $\tau_{\mathrm{DRW}}$ is the damping timescale that characterizes the variability. Then, $\tau_{\mathrm{DRW}}$ should be relevant to characteristic timescales of the accretion disk, such as the thermal timescale. According to the static standard accretion disk theory \citep[hereafter SSD;][]{SSD}, for a given wavelength $\lambda$, if one assumes the corresponding emission-region size $R_{\lambda}$ merits the criteria $k_\mathrm{B}T(R_{\lambda})=hc/\lambda$, the emission-region size $R_{\lambda}/R_\mathrm{S}\propto M_\mathrm{BH}^{-1/3}\dot{m}^{1/3}\lambda^{4/3}$ and the thermal timescale $\tau_{\mathrm{th}}\thicksim \alpha^{-1}\Omega_\mathrm{k}^{-1}\propto \alpha^{-1}\lambda^{2}\dot{M}^{0.5}$, where $k_\mathrm{B}$, $T$, $h$, $c$, $R_\mathrm{S}\equiv2GM_{\mathrm{BH}}/c^2$, $\alpha$, and $\Omega_\mathrm{k}$ are the Boltzmann constant, the effective temperature, the Planck constant, the speed of light, the Schwarzschild radius, the viscosity parameter and the Keplerian angular velocity, respectively; $\dot{m}=\dot{M}/\dot{M}_\mathrm{Edd}=0.1c^2\dot{M}/L_\mathrm{Edd}$ is a dimensionless accretion rate, which is the ratio of the accretion rate $\dot{M}$ to the Eddington accretion rate $\dot{M}_\mathrm{Edd}$, and $L_\mathrm{Edd}$ is the Eddington luminosity. If $\tau_{\mathrm{DRW}}$ is relevant to the thermal timescale, $\tau_{\mathrm{DRW}}$ should also scale as $\lambda^{2}$ or $\dot{M}^{0.5}$. 

Several studies have established empirical relationships between the best-fitting $\tau_{\mathrm{DRW}}$ and $\lambda$ and the physical properties of AGNs. \citet{MacLeod2010} obtained $\tau_{\mathrm{DRW}}\propto \lambda^{0.17}$, $\tau_{\mathrm{DRW}}\propto \dot{M}^{-0.075}$, and $\tau_{\mathrm{DRW}}\propto M_{\mathrm{BH}}^{0.21}$ based on 10-year variability data for 9000 quasars in SSDS Stripe 82; \citet{Suberlak2021} obtained $\tau_{\mathrm{DRW}}\propto \dot{M}^{-0.088}$ and $\tau_{\mathrm{DRW}}\propto M_{\mathrm{BH}}^{0.14}$ based on 15-year variability data for 9000 quasars in SSDS Stripe 82; \citet{Burke2021} obtained $\tau_{\mathrm{DRW}}\propto M_{\mathrm{BH}}^{0.38}$ and $\tau_{\mathrm{DRW}}\propto \dot{M}^{0.33}$ based on the variability data of 67 AGNs, and they even proposed to use the $\tau_{\mathrm{DRW}}-M_{\mathrm{BH}}$ relation to estimate SMBH masses; very recently, \citet{Stone2022} (hereafter S22; also see \citealt{Stone2023} for erratum) obtained $\tau_{\mathrm{DRW}}\propto \lambda^{0.20}$ based on 20-year variability data for 190 quasars in SSDS Stripe 82 (Z. Stone, private communication). These observations demonstrate a weak dependence between the best-fitting $\tau_{\mathrm{DRW}}$ and $\lambda$ and $\dot{M}$ \citep[but see][]{Sun2018, Arevalo2023}. One possible explanation is that the relationship between the damping timescale and the thermal timescale is not a simple linear one. In summary, these observations seem to disfavor the static SSD model strongly. 

It is worth pointing out that when using the DRW model to fit light curves, the durations of light curves (referred to as the baselines) must be much longer than the intrinsic damping timescale; otherwise, the damping timescale will be significantly underestimated. \citet{Kozlowski2017} pointed out that if the intrinsic damping timescale is larger than 10\% of the baseline, the low-frequency white noise cannot be identified accurately, leading to an underestimation of the damping timescale. \citet{Suberlak2021} followed the simulation methodologies of \citet{Kozlowski2017} and revisited the criteria for baselines to obtain unbiased damping timescales; they claim that the best-fitting damping timescale is an unbiased estimator of the intrinsic damping timescale if the best-fitting one is less than 20\% of the baseline. Recently, \citet{Kozlowski2021} stressed that the baseline should be at least 30 times the intrinsic damping timescale. \citet{Hu2023} further suggested that the deviation from the best-fitting damping timescale to the intrinsic value also depends upon the statistical assumptions, including the priors for the DRW parameters and the estimators of the best-fitting damping timescale. In observational studies, it is often argued that if the best-fitting damping timescale is less than 10\% (or 20\%) of the baseline, the best-fitting damping timescale is unbiased \citep[e.g.,][]{Suberlak2021, Burke2021}. \textit{We stress that this criteria is incorrect.} Even if the intrinsic damping timescale is larger than 10\% of the baseline, the best-fitting damping timescale can still be much smaller than 10\% of the baseline in some random realizations of AGN variability (especially for observations with poor cadences). \textit{To verify whether the best-fitting damping timescale is unbiased, one must know the intrinsic damping timescale!} 

S22 studied a sample of 190 quasars with a 20-year baseline in the observed frame using the DRW model. There are only 27 sources that merit the criteria of having a best-fitting damping timescale of less than $20\%$ of the baseline. In addition, the best-fitting damping timescale also has large uncertainties. In our opinion, for the 27 sources, it is still unclear whether their intrinsic damping timescales are less than $20\%$ of the baseline. The same argument also holds for other observational studies \citep[e.g.,][]{Suberlak2021, Burke2021}. In summary, current observational studies of damping timescales may still be seriously limited by the baseline. 

Theoretically, the AGN variability study should not be based on the static SSD model. Instead, time-dependent physical models should be proposed to explain AGN variability. The Corona Heated Accretion disc Reprocessing \citep[CHAR;][]{Sun2020} model established a new connection between disk fluctuations and observed variability; this model can reproduce the SDSS quasar variability \citep[][]{Sun2020-var, Sun2023} and the larger-than-expected UV/optical time lags \citep[][]{LiT2021}. In this model, the black hole accretion disk and the corona are coupled by magnetic fields. When the magnetic field in the corona perturbates, not only does the X-ray luminosity of the corona changes, but also the heating rate of the accretion disk fluctuates. Thus, the temperature of the accretion disk and the UV/optical luminosity vary on the thermal timescales. The CHAR model takes into account the time-dependent evolution of the SSD model to describe the temperature fluctuations, albeit it can be extended to include other disk models. The CHAR model may be effective for understanding the observational results of AGN variability. Is the CHAR model able to account for the observed damping timescales? 

With future wide-field time-domain surveys, e.g., the Wide Field Survey Telescope \citep[WFST;][]{Wang2023-WFST} and the Legacy Survey of Space and Time \citep[LSST;][]{Ivezic2019}, a large amount of AGN data covering multi-band time domains will be available. Extracting information beneficial to AGN studies from a large amount of variability data is crucial. In the future, longer AGN variability data will provide more accurate damping timescales. Hence, it is of great importance to understand the physical nature of the damping timescale.

The main objectives of this paper are to test the CHAR model with S22 observations, to explain the physical nature of the damping timescale in the DRW model, and to propose a new relation between the intrinsic damping timescale and the AGN properties.  

The manuscript is organized as follows. In Section \ref{sec: Consistency}, the CHAR model is compared with the sample of S22; in Section \ref{sec: Interpretation}, we offer new relationships between the intrinsic damping timescales and AGN properties in the CHAR model simulations; in Section \ref{sec: PSD at different radii}, we study the PSD shapes; and Section \ref{sec: Summary} summarizes the main conclusions. 

\section{Consistency between the CHAR Model and Real Observations} \label{sec: Consistency}
In this section, we use the CHAR model to reproduce the DRW fitting results of S22. In Section \ref{subsec: step}, we configure the parameters of the CHAR model. In Sections \ref{subsec: wavelength} and \ref{subsec: PSD}, we use the CHAR model to reproduce the damping timescale dependence on wavelength and the PSDs of the S22 sample. 

\subsection{Setting the parameters for the CHAR model}\label{subsec: step}
The CHAR model uses the static SSD as the initial conditions. Hence, this model requires only three parameters to simulate the light curves: the dimensionless viscosity parameter $\alpha$, the black hole mass $M_{\mathrm{BH}}$, and the dimensionless accretion ratio $\dot{m}(=L_{\mathrm{bol}}/((1+k)L_{\mathrm{Edd}}))$\footnote{Where $k$ is the ratio of the power of magnetic fluctuations in the corona, $Q_{\mathrm{mc}}^+$, to the dissipation rate of the disk turbulent magnetic power, $Q_{\mathrm{vis}}^+$. The value of $k$ does not affect the results, and for ease of computation, $k = 1/3$. For details, see \citet{Sun2020}.}. For all simulations in this paper, we adopt $\alpha = 0.4$ \citep[][]{King2007, Sun2023}. For the 190 quasars in the S22 sample, we use their $M_{\mathrm{BH}}$ and the bolometric luminosity $L_{\mathrm{bol}}$ as the input parameters of the CHAR model (the base parameterization). The black-hole mass is estimated via the single-epoch virial mass estimators, which have substantial uncertainties \citep[$\sim 0.5$ dex; for a review, see, e.g., ][]{Shen2013}. We, therefore, also alter the black hole masses by $0.5$ dex but leave other parameters (e.g., $L_{\mathrm{bol}}$) unchanged, and the resulting damping timescales and their relation to wavelengths are almost unaltered. This is because the damping timescales of the CHAR model are independent of $M_{\mathrm{BH}}$ for fixing $L_{\mathrm{bol}}$ (see Section \ref{subsec: Mbh&mdot}). There is an uncertainty of $0.2-0.3\ \mathrm{dex}$ in the observationally determined $L_{\mathrm{bol}}$ \citep[e.g.,][]{Netzer2019}. To assess the luminosity uncertainties, we consider an extreme case, in which $L_{\mathrm{bol}}$ in the S22 sample are systematically reduced by $ 0.2\ \mathrm{dex}$, leaving the other parameters unchanged (the ``faint" parameterization). \textit{In summary, there is no free parameter in the simulations.} 

S22 uses the $g$, $r$, and $i$ light curves. To facilitate calculations and comparisons, we use the CHAR model to calculate the observed frame (according to each source's redshift) $4500-5500\ \textrm{\AA}$, $5500-6500\ \textrm{\AA}$, and $7000-8000\ \textrm{\AA}$ emission to represent the $g$, $r$, and $i$ bands, respectively. 

\subsection{Dependence of \texorpdfstring{$\tau_{\mathrm{DRW}}$}{} on wavelength}\label{subsec: wavelength}
The baseline of the light curves simulated by the CHAR model is 20 years in the observed frame (i.e., identical to S22). We adopt two sampling patterns for the light-curve simulations: uniform sampling with a cadence of 10 days (hereafter the uniform sampling) and irregular sampling with realistic cadences (hereafter the real sampling). For each quasar with the base and ``faint" parameterizations, we use the CHAR model to generate light curves, encompassing both uniform and real sampling. We fit the light curves with the DRW model and derive the two DRW parameters (i.e., the damping timescale and the variability amplitude) using the \texttt{taufit} code of \citet{Burke2021}\footnote{\url{https://github.com/burke86/taufit}} which is based on the \texttt{celerite} Gaussian-process package \citep{celerite}. The \texttt{celerite} package fits light curves to a given kernel function using the Gaussian Process regression. The DRW kernel function in the \texttt{taufit} code is
\begin{equation}
    k(t_{ij}) = 2\sigma_\mathrm{DRW}^2 e ^{t_{ij}/\tau_\mathrm{DRW}}+\sigma_i^2\delta_{ij},
\end{equation}
where $t_{ij}=\vert t_i-t_j \vert$ is the time interval between two measurements in the light curve, $2\sigma_\mathrm{DRW}^2$ is the long-term variance of variability, $\tau_\mathrm{DRW}$ is the damping timescale, and $\sigma_i^2\delta_{ij}$ denotes an excess white noise term from the measurement errors, with $\sigma_i$ being the excess white noise amplitude and $\delta_{ij}$ being the Kronecker $\delta$ function. The \texttt{taufit} code utilizes the Markov Chain Monte Carlo (MCMC) code of \texttt{emcee} \citep{emcee} with uniform priors to obtain the posterior distributions for the DRW parameters. We take the posterior medians as the best-fitting parameters and the $16^\mathrm{th}$ to $84^\mathrm{th}$ percentiles of the posterior as the $1\sigma$ uncertainties for the measured parameters. The simulation process is repeated $1, 000$ times. Then, we take the medians and $16^\mathrm{th}$ to $84^\mathrm{th}$ percentiles of the one thousand best-fitting damping timescales obtained from the simulations, and the results are shown in Table \ref{tab:tau}. All simulation results are consistent with S22 within $1\sigma$ uncertainties. The sampling slightly affects the best-fitting $\tau_{\mathrm{DRW}}$, and the effect increases with wavelengths. 

\begin{deluxetable*}{cccccc}
\tablenum{1}
\tablecaption{The best-fitting logarithmic $\tau_{\mathrm{DRW}}$ of S22 observations and the CAHR model simulations. \label{tab:tau}}
\tablewidth{0pt}
\tablehead{
\colhead{Bands} & \colhead{Observations} & \multicolumn{2}{c}{Uniform sampling} & \multicolumn{2}{c}{Real sampling} \\
\cline{3-4}
\cline{5-6}
\colhead{} & \colhead{} & \colhead{base} & \colhead{``faint''} & \colhead{base} & \colhead{``faint''}
} 
\colnumbers
\startdata
{$4500-5500\ \textrm{\AA}$ ($g$ band)}&$2.88^{+0.56}_{-0.28}$ & $3.20^{+0.17}_{-0.29}$&$3.13^{+0.18}_{-0.30}$ &$3.19^{+0.11}_{-0.15}$&$3.15^{+0.12}_{-0.18}$ \\
{$5500-6500\ \textrm{\AA}$ ($r$ band)}&$3.05^{+0.61}_{-0.34}$  & $3.33^{+0.16}_{-0.26}$& $3.25^{+0.17}_{-0.30}$& $3.23^{+0.10}_{-0.13}$&$3.20^{+0.11}_{-0.14}$\\
{$7000-8000\ \textrm{\AA}$ ($i$ band)}&$3.11^{+0.58}_{-0.35}$  & $3.47^{+0.16}_{-0.25}$& $3.40^{+0.16}_{-0.25}$& $3.30^{+0.08}_{-0.10}$&$3.27^{+0.09}_{-0.12}$\\
\enddata
\tablecomments{Column (1) represents the wavelength ranges in the observed frame; Column (2) is the results obtained by S22; Columns (3) and (4) are the CHAR model simulation results with uniform sampling for the base and ``faint" parameterizations, respectively; Columns (5) and (6) are the CHAR model simulation results with real sampling.}
\end{deluxetable*}

\begin{figure*}
    \centering
    \includegraphics{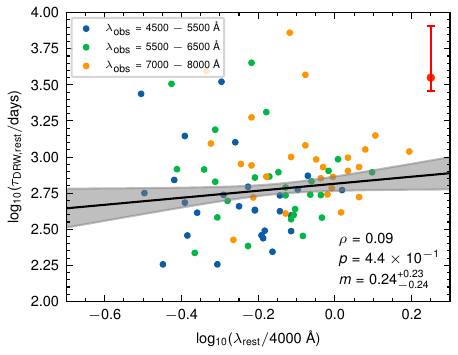}
    \includegraphics{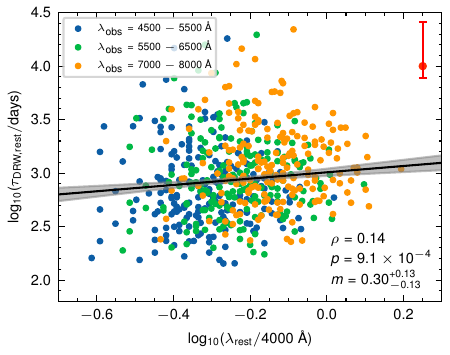}
    \caption{A typical realization of the dependence of the best-fitting $\tau_{\mathrm{DRW}}$ on wavelength obtained from the CHAR model. $M_{\mathrm{BH}}$, $L_{\mathrm{bol}}$, and redshifts are taken from the S22 sample. In each panel, the red error bar in the upper right corner shows the median $1\sigma$ uncertainty for all best-fitting $\tau_{\mathrm{DRW}}$. The Spearman's rank correlation coefficient ($\rho$), $p$-value, and the best-fitting slope ($m$) are shown in the lower right corner. The left panel is for the same subsample in S22 (i.e., with 27 quasars) and the correlation is statistically insignificant. The right panel is for the full sample of 190 quasars in S22; in this sample, the correlation is statistically significant. The solid lines and the shaded areas are the best-fitting lines and $1\sigma$ uncertainties.}
    \label{fig: wavelength}
\end{figure*}

We then investigate the dependence of the best-fitting $\tau_\mathrm{DRW}$ on wavelength obtained from the CHAR model and compare them with the results in S22. Previous studies have shown that the best-fitting $\tau_\mathrm{DRW}$ is significantly underestimated if the baseline is not longer than ten (or five) times the intrinsic damping timescale \citep[][]{Kozlowski2017, Suberlak2021, Hu2023}. S22 selects a subsample with the best-fitting damping timescales less than 20\% of the baseline containing 27 sources. As mentioned in Section~\ref{sec: intro}, their source selection criteria cannot eliminate the effects of baseline inadequacy. The real observations in S22 yield $\tau_\mathrm{DRW,S22}\propto \lambda^{0.20\pm 0.20}$ for the subsample and $\tau_\mathrm{DRW,S22}\propto \lambda^{0.30\pm0.13}$ for the full sample.\footnote{The slope reported here is slightly smaller (but statistically consistent within $1\sigma$) than \cite{Stone2023}. This is because the fitting code used by \cite{Stone2023} has a minor bug (Z. Stone, private communication).} Figure \ref{fig: wavelength} shows a typical realization of the dependence of best-fitting rest-frame $\tau_{\mathrm{DRW,CHAR}}$ on wavelength obtained from the CHAR model: the left panel includes only the subsample in S22, while the right panel shows the result for their full sample. We fit the $\tau_\mathrm{DRW,CHAR}-\lambda_\mathrm{rest}$ relation with $\mathrm{log_{10}}\tau_\mathrm{DRW,CHAR}=m\mathrm{log_{10}}\lambda_\mathrm{rest} + n$. The posterior distributions of $m$ and $n$ are obtained by the MCMC code \texttt{emcee} \citep[][]{emcee} with uniform priors, and the logarithmic likelihood function is $\ln \mathcal{L}=-\!0.5\sum \left\{(\log_{10} \tau_{\mathrm{DRW, CHAR}}\!-\!(m\log_{10} \lambda_{\mathrm{rest}}+n))^2/\sigma_\mathrm{CHAR}^2 \notag\right. \\ \left. + \ln{\sigma_\mathrm{CHAR}^2}\right\}$, where $\sigma_\mathrm{CHAR}$ is the $1\sigma$ uncertainty of the best-fitting $\tau_{\mathrm{DRW, CHAR}}$. The best-fitting results for $m$ and $n$ are taken as the posterior medians, and their $1\sigma$ uncertainties are taken as $16^\mathrm{th}$ to $84^\mathrm{th}$ percentiles of the posterior distribution. 

For different sampling methods and parameterizations, we obtain the probabilities of having the CHAR-model slope $m$ agree with S22 within $1\sigma$ uncertainty through one thousand repeated CHAR model simulations, and the results are shown in Table \ref{tab:slope}.
Thus, we cannot statistically reject the hypothesis that the dependence of the best-fitting $\tau_{\mathrm{DRW}}$ on wavelength obtained by the CHAR model is statistically consistent with S22.

\begin{deluxetable}{ccccc}
\tablenum{2}
\tablecaption{Probabilities of reproducing observations \label{tab:slope}}
\tablewidth{0pt}
\tablehead{
\colhead{Sample} & \multicolumn{2}{c}{Uniform sampling} & \multicolumn{2}{c}{Real sampling} \\
\cline{2-3}
\cline{4-5}
\colhead{} & \colhead{base} & \colhead{``faint''} & \colhead{base} & \colhead{``faint''}
} 
\startdata
Subsample&62.0\% & 77.4\%&75.0\% &77.9\%\\
Full sample&40.3\%  &76.8\%& 91.2\%& 84.4\%\\
\enddata
\tablecomments{The table values are the probabilities of the simulated slope $m$ matching the observed value (within the $1\sigma$ confidence interval).}
\end{deluxetable}

Real observations show that the best-fitting $\tau_{\mathrm{DRW}}$ exhibits a weak dependence on wavelength.
One possible reason is that the baseline is not long enough, leading to an underestimation of the intrinsic $\tau_{\mathrm{DRW}}$ (see Section \ref{subsec:simulation steps} for details). We stress that, as we mentioned in Section \ref{sec: intro}, this bias still exists even if one only selects sources with the best-fitting damping timescale less than $10\%$ (or $20\%$) of the baseline. Indeed, if we increase the simulation baseline, the best-fitting damping timescale also increases, and the timescale-wavelength relation has a larger slope. Another possible reason is that the observed $\tau_{\mathrm{DRW}}$ is the average of thermal timescales at different radii of the accretion disk, hence the relationship between $\tau_{\mathrm{DRW}}$ and wavelength may not be very simple (see Section \ref{subsec:R_var} for details).

\subsection{Power spectral density} \label{subsec: PSD}
\begin{figure*}
    \centering
    \includegraphics[width=21cm,height=7cm,keepaspectratio]{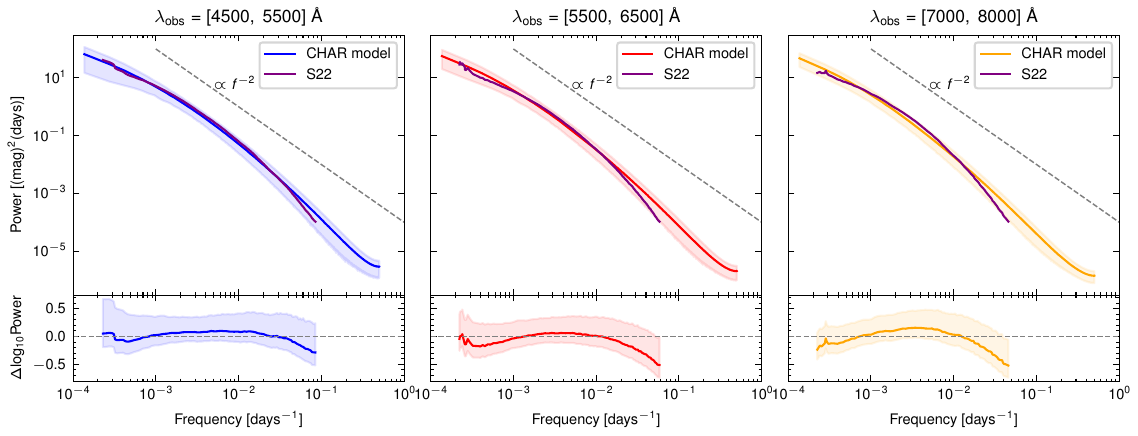}
    \caption{Comparisons of the CHAR model and real observations of rest-frame ensemble PSDs for the full sample in different bands. The purple curves are the real observations from S22. The blue, red, and yellow curves and shaded areas are the simulation results of the CHAR model and their $1\sigma$ confidence intervals. The dashed grey lines represent the $f^{-2}$ power-law function. The lower panels are the differences between the CHAR model and the real observations on a logarithmic scale. The ensemble PSDs of the CHAR model are consistent with those of the S22 observations within the $1\sigma$ confidence intervals.}
    \label{fig: compare PSD}
\end{figure*}

Power spectral density is one of the most essential tools for studying AGN variability. We use the CHAR model to generate light curves for the full sample in S22 with a 20-year baseline in the observed frame and a cadence of 1 day. Because the light curves generated by the CHAR model are uniformly sampled, we use the Fast Fourier Transform (FFT) to obtain the model ensemble PSDs for the full sample. We repeat the above process two hundred times and then take the average of these simulations to obtain the PSDs. Whereas the light curves of the observations are not uniformly sampled, S22 used the continuous autoregressive with moving-average \citep[CARMA;][]{kelly2014} model, which is a high-order model, to obtain the ensemble PSDs. Figure \ref{fig: compare PSD} compares the CHAR model with real observations of rest-frame ensemble PSDs for the full sample in different bands. The ensemble PSDs of the CHAR model are consistent with S22 observations within the $1\sigma$ confidence interval. At high frequencies, the ensemble PSDs of both the S22 observations and the CHAR model simulations are steeper than the $f^{-2}$ power-law function, suggesting that the DRW model overestimates the variability on short timescales, which is consistent with previous Kepler studies \citep{Mushotzky2011, Kasliwal2015, Smith2018}.

\begin{figure}
    \centering
    \includegraphics{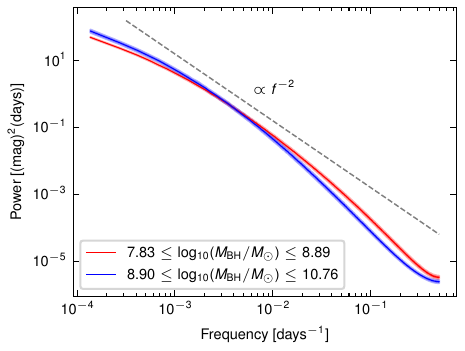}
    \caption{The ensemble PSDs of the CHAR model at the $\lambda_{\mathrm{obs}}=4500-5500\ \textrm{\AA}$ band for two $M_{\mathrm{BH}}$ subsamples defined in S22. The dashed grey line represents the $f^{-2}$ power-law function. The high-mass sample tends to have a steeper ensemble PSD than the low-mass one. }
    \label{fig: ensPSD by mbh}
\end{figure}

\begin{figure*}
    \centering
    \includegraphics[width=17cm,height=17cm,keepaspectratio]{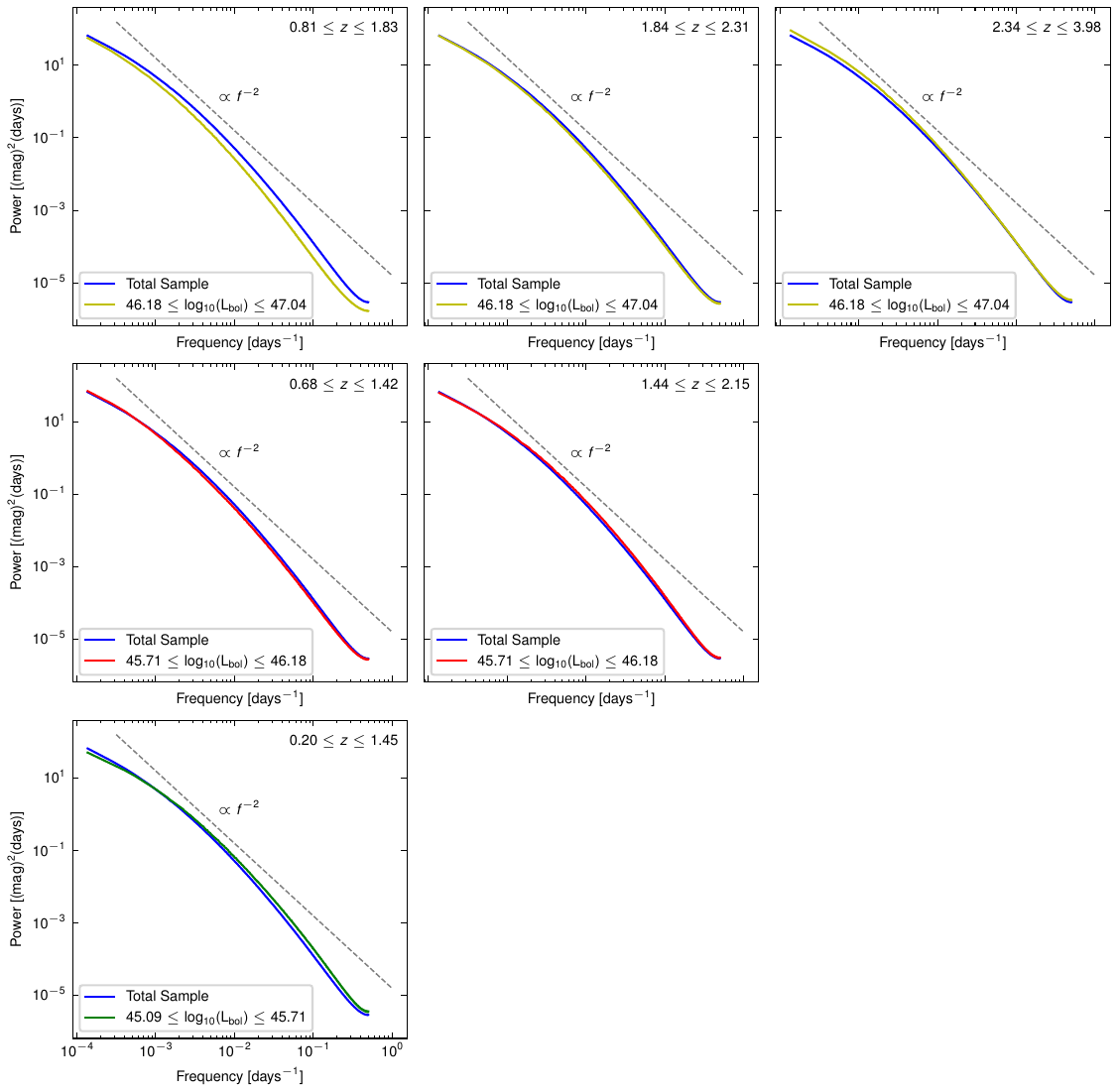}
    \caption{The ensemble PSDs of the CHAR model in the $\lambda_{\mathrm{obs}}=4500-5500\ \textrm{\AA}$ band for different $L_{\mathrm{bol}}$ and redshift groups. The grouping strategy is the same as Figure 16 in S22. The dashed grey lines represent the $f^{-2}$ power-law function. The dependence of the ensemble PSDs upon $L_{\mathrm{bol}}$ and redshift is similar to that of S22.}
    \label{fig: ensPSD by L}
\end{figure*}

As in S22, we also study the model ensemble PSDs after grouping the full sample by different methods. Figures \ref{fig: ensPSD by mbh} and \ref{fig: ensPSD by L} show the ensemble PSDs of the subsamples grouped by $M_{\mathrm{BH}}$ and the subsamples grouped by $L_{\mathrm{bol}}$ and redshift in $\lambda_{\mathrm{obs}}=4500-5500\ \textrm{\AA}$ band, respectively. The grouping strategies are the same as Figures 16 and 18 of S22. The model ensemble PSDs of the subsamples exhibit similar behaviors to those of S22: the high-frequency breaks of quasars with smaller SMBHs or lower luminosities tend to occur on shorter timescales and vice versa. 

In summary, our studies above suggest that the CHAR model can reproduce the ensemble PSDs of the sample in S22. We then can use the CHAR model to predict the intrinsic damping timescale for other AGNs and its dependence upon AGN properties after properly eliminating the biases due to the limited baseline. 

In the observational point of view, the DRW (a.k.a., the CARMA(1, 0) model) modeling is still an efficient way to understand AGN UV/optical variability. First, the real light curves are often very sparse. Several sophisticated statistical models are proposed to fit AGN UV/optical light curves, e.g., the damped harmonic oscillator \citep[DHO, a.k.a., CARMA(2, 1);][]{Moreno2019} model and other high-order CARMA models \citep{kelly2014}. Compared with the DRW model, these sophisticated statistical models have more free parameters. For most AGN UV/optical light curves, these free parameters cannot be simultaneously well constrained. Second, the damping timescale of the DRW model is related to the timescales of the DHO model \citep[see the lower-right panel of Figure 14 in][]{Yu2022}. \citet{Kasliwal2017} compared the PSDs of the DRW and the DHO models and found no difference between the two PSDs on timescales longer than $\sim 10$ days. Hence, while the ensemble PSD analysis suggests that AGN UV/optical variability is not consistent with the DRW model (Figures~\ref{fig: compare PSD}, \ref{fig: ensPSD by mbh} and \ref{fig: ensPSD by L}), one can still use this model to measure the long breaking timescales. We stress that, as pointed out by \cite{Vio1992}, there is no one-to-one relationship between statistical models and intrinsic dynamics. Ideally, one should directly fit real AGN UV/optical light curves with physical models that have been tested (e.g., the CHAR model), which is beyond the scope of this manuscript.

\section{Physical Interpretation for the Damping Timescale} \label{sec: Interpretation}
We now discuss the intrinsic damping timescale in the CHAR model. The simulation steps are introduced in Section \ref{subsec:simulation steps}; the relationship between the intrinsic damping timescale and $M_{\mathrm{BH}}$, $\dot{m}$ and rest-frame wavelength $\lambda_\mathrm{rest}$ is given in Section \ref{subsec: Mbh&mdot}; the relationships between the intrinsic damping timescale and a series of physical properties are presented in Section \ref{subsec:R_var}.

\subsection{Simulation steps} \label{subsec:simulation steps}
The simulation parameter settings and key points are as follows:
\begin{enumerate}
\item The simulation parameters $M_{\mathrm{BH}}$ and $\dot{m}$ used for the CHAR model are shown in Table \ref{tab:parameters1}, including 27 cases with the bolometric luminosity larger than $10^{44}\ \mathrm{erg\ s^{-1}}$. 
\item The simulation bands are shown in Table \ref{tab:parameters2} which cover rest-frame UV-to-optical wavelengths.
\item The light curves of the integrated thermal emission from the whole disk with rest-frame 20-year and 40-year baselines are simulated using the CHAR model with a cadence of 10 days. The light curves are then fitted using the DRW model, and the best-fitting damping timescale is denoted as $\tau_{\mathrm{DRW, disk}}$. The simulation is repeated unless at least one hundred sets of light curves are obtained for each case in Table \ref{tab:parameters1}. 
\item For the rest-frame 20-year baseline, the light curves at different radii of the accretion disk are also fitted using the DRW model, and their best-fitting damping timescales are denoted as $\tau_{\mathrm{DRW, radius}}$.
\end{enumerate}

\begin{deluxetable}{cccccccc}
\tablenum{3}
\tablecaption{Model parameters for the CHAR model simulation\label{tab:parameters1}}
\tablewidth{0pt}
\tablehead{
\colhead{Number} & \colhead{$\mathrm{log_{10}}(M_{\mathrm{BH}}/M_{\odot})$} & \colhead{$\dot{m}$} & \colhead{$\mathrm{log_{10}}(L_{\mathrm{bol}}/[\mathrm{erg\ s^{-1}}])$}
} 
\startdata
1* & 8.0 & 0.01 & 44.24 \\
2* & 8.5 & 0.01 & 44.74 \\
3 & 9.0 & 0.01 & 45.24 \\
4* & 7.5 & 0.05 & 44.44 \\
5* & 8.0 & 0.05 & 44.94 \\
6 & 8.5 & 0.05 & 45.44  \\
7 & 9.0 & 0.05 & 45.94  \\
8* & 7.0 & 0.1 & 44.24  \\
9* & 7.5 & 0.1 & 44.74 \\
10 & 8.0 & 0.1 & 45.24 \\
11 & 8.5 & 0.1 & 45.74 \\
12 & 9.0 & 0.1 & 46.24 \\
13* & 7.0 & 0.15 & 44.41 \\
14* & 7.5 & 0.15 & 44.91 \\
15 & 8.0 & 0.15 & 45.41 \\
16 & 8.5 & 0.15 & 45.91 \\
17 & 9.0 & 0.15 & 46.41 \\
18* & 7.0 & 0.2 & 44.54 \\ 
19* & 7.5 & 0.2 & 45.04 \\ 
20 & 8.0 & 0.2 & 45.54 \\
21 & 8.5 & 0.2 & 46.04 \\ 
22 & 9.0 & 0.2 & 46.54 \\
23* & 7.0 & 0.5 & 44.94 \\
24 & 7.5 & 0.5 & 45.44 \\
25 & 8.0 & 0.5 & 45.94 \\
26 & 8.5 & 0.5 & 46.44 \\
27 & 9.0 & 0.5 & 46.94 \\
\enddata
\tablecomments{Columns (1) shows the case number, those with * are cases with $L_{\mathrm{bol}}<10^{45.1}\ \mathrm{erg\ s^{-1}}$; columns (2) represents the logarithmic black-hole mass; columns (3) indicates the dimensionless accretion ratio $\dot{m}$; and columns (4) shows the logarithmic bolometric luminosity.}
\end{deluxetable}

\begin{deluxetable}{cc}
\tablenum{4}
\tablecaption{Simulation bands\label{tab:parameters2}}
\tablewidth{0pt}
\tablehead{
\colhead{Band id. } & \colhead{Wavelength ranges}\\ \colhead{} & \colhead{[$\textrm{\AA}$]}
} 
\startdata
1 & 1500--2500\\
2 & 2500--3500\\
3 & 3500--4500\\
4 & 4500--5500\\
5 & 5500--6500\\
6 & 7000--8000\\
\enddata
\end{deluxetable}

\begin{figure*}
    \centering
    \includegraphics{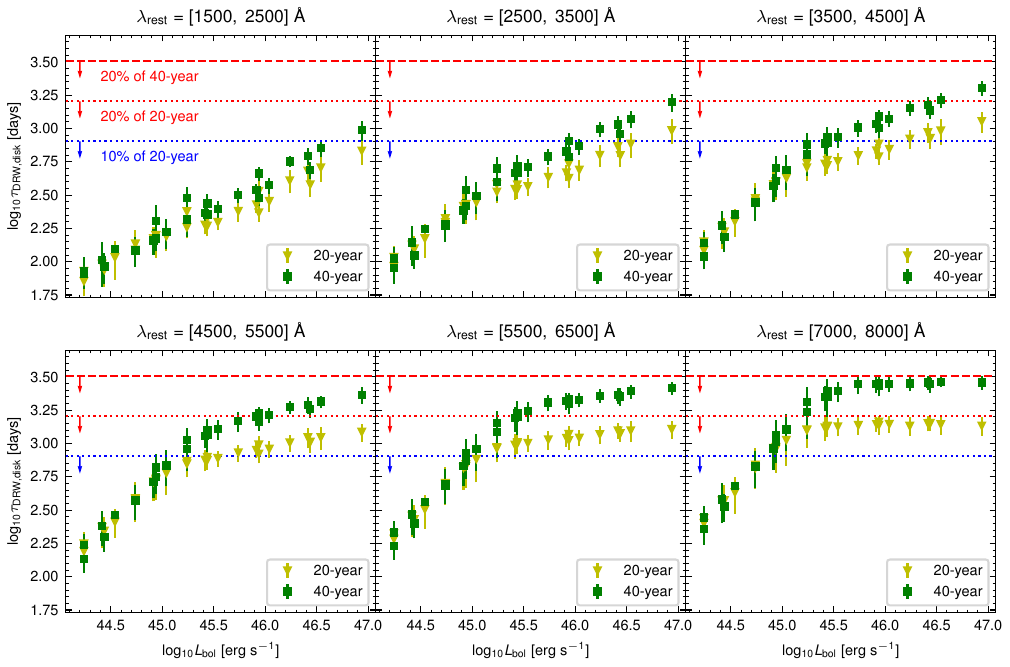}
    \caption{The relationship between the best-fitting damping timescale of the integrated disk light curve $\tau_{\mathrm{DRW,disk}}$ and $L_{\mathrm{bol}}$. Different panels indicate different bands. The yellow inverted triangles and green squares are the medians of the one hundred simulations of the CHAR model for the 20-year and 40-year baselines, respectively. The error bars are $1\sigma$ uncertainties. The red dotted and red dashed lines represent 20\% of the 20-year and 40-year baselines, respectively. The blue dotted lines represent 10\% of the 20-year baseline. When $L_{\mathrm{bol}}>10^{45.1}\ \mathrm{erg\ s^{-1}}$, $\tau_{\mathrm{DRW,disk}}$ is biased since its values for the 20-year baseline are significantly smaller than the 40-year baseline. }
    \label{fig: tau_disc-Lbol}
\end{figure*}

If the baseline is not longer than ten (or five) times the intrinsic damping timescale, $\tau_{\mathrm{DRW,disk}}$ will be strongly underestimated \citep[e.g.,][]{Kozlowski2017, Suberlak2021, Hu2023}. Under such circumstances, the fitted $\tau_{\mathrm{DRW,disk}}$ should increase with the baseline. Figure \ref{fig: tau_disc-Lbol} shows the relationship between the median $\tau_{\mathrm{DRW,disk}}$ over one hundred simulations and $L_{\mathrm{bol}}$ at different bands, for both the 20-year and 40-year baselines. Note that, here we only consider simulations with $\tau_{\mathrm{DRW, disk}}/\mathrm{baseline}<20\%$ for all bands. At the high bolometric luminosity end, there are clear differences between $\tau_{\mathrm{DRW,disk}}$ obtained for the 20-year baseline and the 40-year baseline, with the results for the 20-year baseline being significantly smaller than the 40-year baseline. The difference is more significant at longer wavelength bands. Hence, even if one only keeps the simulations with the best-fitting damping timescale less than 20\% of the baseline, the intrinsic damping timescale cannot be obtained for the high luminosity cases. Likewise, in real observations, the best-fitting damping timescale is also biased even if it is less than 20\% of the observed baseline. 

As we mentioned in Section \ref{sec: intro}, the requirement for recovering the intrinsic damping timescale is that the intrinsic damping timescale (rather than the best-fitting one) should be less than $10\%$ (or 20\%) of the baseline. The results of \citet{Hu2023} show that if the intrinsic $\tau_{\mathrm{DRW}}$ is 10\% of the baseline, although the statistical expectation of the output best-fitting $\tau_{\mathrm{DRW}}$ is the same as the intrinsic $\tau_{\mathrm{DRW}}$, its $1\sigma$ dispersion is as large as 50\%. If one only requires that the best-fitting $\tau_{\mathrm{DRW}}$ is less than $10\%$ (or $20\%$) of the baseline, the statistical expectation of $\tau_{\mathrm{DRW}}$ can still be significantly biased.  This is because a DRW model with a large intrinsic $\tau_{\mathrm{DRW}}$ (larger than $10\%$ or $20\%$ of the baseline) can accidentally have a small best-fitting $\tau_{\mathrm{DRW}}$ because of statistical fluctuations. 

If $\tau_{\mathrm{DRW,disk}}$ does not increase with baseline, one can regard $\tau_{\mathrm{DRW,disk}}$ as the intrinsic one. Practically speaking, the difference ($\Delta\mathrm{log_{10}}\tau_{\mathrm{DRW,disk}}$) between the 20-year and 40-year baseline results is less than $0.1$ at the longest wavelength range we probed ($[7000,\ 8000]\ \textrm{\AA}$). Thus, for the 20-year baseline, $\tau_{\mathrm{DRW,disk}}$ is reliable (i.e., the best-fitting $\tau_{\mathrm{DRW,disk}}$ equal to the intrinsic value) only when $L_{\mathrm{bol}}<10^{45.1}\ \mathrm{erg\ s^{-1}}$ (cases marked with * in Table \ref{tab:parameters1}). For such sources, the best-fitting $\tau_{\mathrm{DRW,disk}}$ for the 20-year baseline is the same as that for the 40-year baseline. For the $L_{\mathrm{bol}}<10^{45.1}\ \mathrm{erg\ s^{-1}}$ cases, their $\tau_{\mathrm{DRW,disk}}$ is less than 10\% of baseline, which is consistent with \citet{Kozlowski2017} and \citet{Hu2023}. We only take the cases with $L_{\mathrm{bol}}<10^{45.1}\ \mathrm{erg\ s^{-1}}$ in Table \ref{tab:parameters1} for subsequent analysis, but our conclusions are generalizable as long as the baseline is long enough. 

\subsection{\texorpdfstring{The dependencies of $\tau_{\mathrm{DRW,disk}}$ upon $M_{\mathrm{BH}}$, $\dot{m}$, and $\lambda_{{\mathrm{rest}}}$}{}}\label{subsec: Mbh&mdot} 
\begin{figure*}
    \centering
    \includegraphics{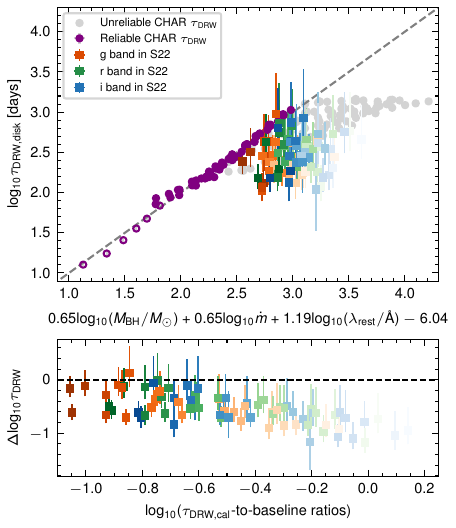}
    \includegraphics{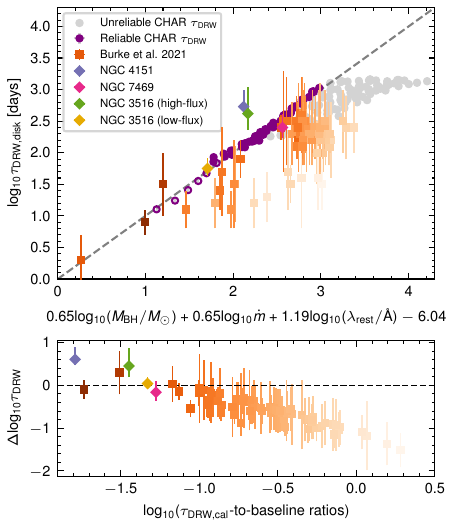}
    \caption{The relationship between $\tau_{\mathrm{DRW, disk}}$ and $M_{\mathrm{BH}}$, $\dot{m}$, and $\lambda_{\mathrm{rest}}$ (see Equation \ref{equ: tau_disc-M-mdot}). 
    In both panels, the purple-filled and grey dots represent the CHAR model prediction for cases in Table \ref{tab:parameters1} with $L_{\mathrm{bol}}<10^{45.1}\ \mathrm{erg\ s^{-1}}$ (whose $\tau_{\mathrm{DRW, disk}}$ is unbiased for the 20-year baseline) and $L_{\mathrm{bol}}>10^{45.1}\ \mathrm{erg\ s^{-1}}$ (whose $\tau_{\mathrm{DRW, disk}}$ is strongly biased for the 20-year baseline), respectively. The purple-open dots are the CHAR model calculations for a low-luminosity case with $M_{\mathrm{BH}}=10^{7.0}\ M_\odot$ and $\dot{m}=0.01$. The gray dashed lines indicate the one-to-one relation. For comparison purposes, we also include real observations. 
    The 27 sources in the left panel are taken from S22, where $\tau_{\mathrm{DRW,obs}}$ is less than 20\% of the baseline. Dots of different colors indicate different bands. In the right panel, the orange dots indicate the 67 sources in  \citet{Burke2021}, and the color diamonds are model results for the cases in Table \ref{tab:sources}. Darker colors indicate smaller $\tau_{\mathrm{DRW,cal}}$-to-baseline ratios in the rest-frame. The lower panels show the relationship between $\Delta\mathrm{log_{10}}{\tau_\mathrm{DRW}}=\mathrm{log_{10}}(\tau_{\mathrm{DRW,obs}}/\tau_{\mathrm{DRW,cal}})$ and $\tau_{\mathrm{DRW,cal}}$-to-baseline ratios in the rest-frame. $\Delta\mathrm{log_{10}}{\tau_\mathrm{DRW}}$ decreases with $\tau_{\mathrm{DRW,cal}}$-to-baseline ratios, which strongly suggests that the observationally-determined damping timescales are significantly underestimated.}
    \label{fig: Stone&Burke}
\end{figure*}

Previous studies have suggested that the best-fitting damping timescale may be related to the SMBH mass or the AGN luminosity and established empirical relationships between them based on observations \citep[e.g.,][]{MacLeod2010, Burke2021, Wang2023}. However, real observations cannot properly eliminate the effects of inadequate baselines. We consider only the cases with $L_{\mathrm{bol}}<10^{45.1}\ \mathrm{erg\ s^{-1}}$ in Table \ref{tab:parameters1} and use the medians of one hundred simulations of $\tau_{\mathrm{DRW,disk}}$ obtained from step 4 in Section \ref{subsec:simulation steps} to establish the relationship between $\tau_{\mathrm{DRW,disk}}$ and $M_{\mathrm{BH}}$, $\dot{m}$, and rest-frame wavelength $\lambda_{\mathrm{rest}}$. $\lambda_{\mathrm{rest}}$ is taken as the medians of the different bands in Table \ref{tab:parameters2}. The fitted equation is
\begin{equation}\label{equ: tau_disc-M-mdot}
    \begin{split}
        \mathrm{log_{10}}\tau_{\mathrm{DRW,disk}}=&a\mathrm{log_{10}}(M_{\mathrm{BH}}/M_{\odot})+b\mathrm{log_{10}}\dot{m} \\
        &+c\mathrm{log_{10}}(\lambda_{\mathrm{rest}}/\textrm{\AA})+d,
    \end{split}
\end{equation}
The MCMC code \texttt{emcee} \citep{emcee} is adopted to sample the posterior distributions of the fitting parameters with the model likelihood and uniform priors. The logarithmic likelihood function is $\ln \mathcal{L}= -0.5\sum \left\{(f_\mathrm{i}-f_\mathrm{model,i})^2/\sigma_\mathrm{i}^2+\ln{\sigma_\mathrm{i}^2} \right\}$, where $f_\mathrm{i}$ and $f_\mathrm{model,i}$, and $\sigma_\mathrm{i}$ are the measured damping timescale, the model timescale, the $1\sigma$ uncertainty of $f_i$, respectively. The best-fitting values for the parameters are taken as the posterior medians, and their $1\sigma$ uncertainties are taken as $16^\mathrm{th}$ to $84^\mathrm{th}$ percentiles of the posterior distribution, as are all subsequent fits. Figure \ref{fig: Stone&Burke} shows the relationship between $\tau_{\mathrm{DRW,disk}}$ and $M_{\mathrm{BH}}$, $\dot{m}$, and $\lambda_{{\mathrm{{\mathrm{rest}}}}}$ (purple-filled dots). The best-fitting parameters and their $1\sigma$ uncertainties are $a\! = \!0.65^{+0.01}_{-0.01}, b\! = \!0.65^{+0.01}_{-0.01}, c\! = \!1.19^{+0.01}_{-0.01}, d\! =\! -\!6.04^{+0.05}_{-0.05}.$ The result demonstrates that $\tau_{\mathrm{DRW,disk}}\propto L_\mathrm{bol}^{0.65}\lambda^{1.19}$, i.e., $\tau_{\mathrm{DRW,disk}}$ is strongly related to the bolometric luminosity $L_{\mathrm{bol}}$ and the rest-frame wavelength $\lambda_{\mathrm{rest}}$, with little or no correlation with $M_\mathrm{BH}$, which ensures the feasibility of using the damping timescale to probe the cosmological time dilation \citep{Lewis2023}. 

For the cases in Table \ref{tab:parameters1} but with $L_{\mathrm{bol}}>10^{45.1}\ \mathrm{erg\ s^{-1}}$ (gray dots in Figure \ref{fig: Stone&Burke}), their best-fitting damping timescales are strongly underestimated for the rest-frame 20-year baseline. Hence, they fall below the relation of Equation \ref{equ: tau_disc-M-mdot}. 

Our relationship (Equation \ref{equ: tau_disc-M-mdot}) also holds for luminosity ranges lower than all cases in Table~\ref{tab:parameters1}. To demonstrate this, we consider a low-luminosity case with $M_{\mathrm{BH}}=10^{7.0}\ M_\odot$ and $\dot{m}=0.01$. We set a cadence of one day rather than ten days because the expected damping timescales (Equation~\ref{equ: tau_disc-M-mdot}) can be as short as ten days. Same as in Section \ref{subsec:simulation steps}, we only consider simulations with $\tau_{\mathrm{DRW, disk}}/\mathrm{baseline}<20\%$ for all bands. The simulations are repeated $100$ times. The purple-open dots in Figure \ref{fig: Stone&Burke} are the median values of the hundred simulations. These damping timescales are in good agreement with Equation \ref{equ: tau_disc-M-mdot}.

In Equation \ref{equ: tau_disc-M-mdot}, $\tau_{\mathrm{DRW,disk}}\propto \lambda^{1.19}$, whereas the relationships between the best-fitting damping timescale and wavelength obtained in both Figure \ref{fig: wavelength} and real observations of S22 are much weaker than this relationship. This is because, for most of the targets in S22, the 20-year baseline in the observed frame does not yield unbiased damping timescales. For the subsample in S22, the selection criterion is that the best-fitting damping timescale rather than the intrinsic one is less than $20\%$ of the baseline; hence, the best-fitting damping timescales in the subsample are also biased to smaller values. The bias should increase with wavelength if the intrinsic damping timescale positively correlates with the wavelength. As a result, the dependence of best-fitting damping timescale and wavelength obtained by S22 is weaker than Equation \ref{equ: tau_disc-M-mdot}. \cite{Lewis2023} uses the damping timescales of the S22 sample to probe the cosmological time dilation. Their conclusion might also be affected by the same bias. 

The Equation \ref{equ: tau_disc-M-mdot} can predict a given AGN's intrinsic damping timescale and justify whether or not the best-fitting damping timescales in observational studies are biased. Figure \ref{fig: Stone&Burke} compares the best-fitting damping timescales from real observations (hereafter $\tau_{\mathrm{DRW,obs}}$) with Equation \ref{equ: tau_disc-M-mdot} (hereafter $\tau_{\mathrm{DRW,cal}}$). The left panel presents the subsample of S22 with the best-fitting damping timescales less than 20\% of the baseline. For almost all sources, the $\tau_{\mathrm{DRW,cal}}$-to-baseline ratio is significantly larger than $10\%$, and the best-fitting damping timescales should be strongly biased. Hence, $\tau_{\mathrm{DRW,obs}}$ is always smaller than $\tau_{\mathrm{DRW,cal}}$, just like the gray dots in Figure \ref{fig: Stone&Burke}. Moreover, targets with smaller $\Delta\mathrm{log_{10}}{\tau_\mathrm{DRW}}=\mathrm{log_{10}}(\tau_{\mathrm{DRW,obs}}/\tau_{\mathrm{DRW,cal}})$ values tend to have larger $\tau_{\mathrm{DRW,cal}}$-to-baseline ratios. This anti-correlation again strongly supports the idea that the best-fitting damping timescales in S22 are probably underestimated. The right panel shows the 67 AGNs in \citet{Burke2021}. Again, we find that, for sources with large $\tau_{\mathrm{DRW,cal}}$-to-baseline ratios (i.e., $>10\%$), their $\tau_{\mathrm{DRW,obs}}$ are lower than the predictions from Equation \ref{equ: tau_disc-M-mdot}. Interestingly, five sources\footnote{The five sources are NGC 4395, NGC 5548, SDSS J025007.03+002525.3, SDSS J153425.58+040806.7, and SDSS J160531.85+174826.3.} in \citet{Burke2021} have small $\tau_{\mathrm{DRW,cal}}$-to-baseline ratios (i.e., $<10\%$), and their best-fitting damping timescales agree well with Equation \ref{equ: tau_disc-M-mdot}. \cite{Burke2021} combined the timescale measurements from AGNs and white dwarfs. The white dwarfs have short damping timescales of $\lesssim 0.01$ day and can be unbiasedly measured. Then, they found that $\tau_{\mathrm{DRW}}\propto M_\mathrm{BH}^{0.5}$. Given that the sources in their sample roughly have similar $\dot{m}$, their result also suggests that $\tau_{\mathrm{DRW}}\propto L_\mathrm{bol}^{0.5}$, which is close to our Equation \ref{equ: tau_disc-M-mdot}.

To further test Equation \ref{equ: tau_disc-M-mdot}, we measure the damping timescale for three local AGNs listed in Table \ref{tab:sources}. These sources have relatively long baselines and small luminosities or black-hole masses. According to Equation \ref{equ: tau_disc-M-mdot}, these sources should have small intrinsic damping timescales and can be unbiasedly measured. Their light curves are obtained from literature as indicated in Table~\ref{tab:sources}. We again use \texttt{taufit} to measure their best-fitting damping timescales and $1\sigma$ uncertainties (diamonds in the right panel of Figure \ref{fig: Stone&Burke}). The results are again roughly consistent with Equation \ref{equ: tau_disc-M-mdot}. Interestingly, two sources with relatively large deviations, i.e., NGC 4151 and NGC 3516, are both changing-look AGNs. For NGC 3516, we separately measured their damping timescales in the high-flux state (from 1996 to 2007) and low-flux state (from 2018 to 2021). In the high-flux state, the measured damping timescale is consistent with our relation within $2\sigma$; in the low-flux state, the measured damping timescale is almost identical to $\tau_{\mathrm{DRW,cal}}$. Hence, our results suggest that the thermal structure of changing-look AGN's accretion disks changes as the line appears or disappears.

\begin{deluxetable*}{ccccccccccc}
\tablenum{5}
\tablecaption{Additional low-luminosity targets with long light curves. \label{tab:sources}}
\tabletypesize{\scriptsize}
\tablehead{
\colhead{Object} & \colhead{$z$} & \colhead{$M_{\mathrm{BH}}$} & \colhead{$L_{\mathrm{bol}}$} & \colhead{$\lambda_\mathrm{rest}$} & \colhead{Baseline} & \colhead{$\mathrm{log_{10}}(\tau_\mathrm{DRW,obs}/[\mathrm{days}])$} & \colhead{$\frac{\tau_\mathrm{DRW,obs}}{\mathrm{baseline}}$}& \colhead{$\mathrm{log_{10}}(\tau_\mathrm{DRW,cal}/[\mathrm{days}])$} & \colhead{$\frac{\tau_\mathrm{DRW,cal}}{\mathrm{baseline}}$} & \colhead{Ref.} \\
\colhead{}&\colhead{}&\colhead{[$10^7M_\odot$]}&\colhead{[$10^{44}\ \mathrm{erg\ s^{-1}}$]}&\colhead{[$\textrm{\AA}$]}&\colhead{[$\mathrm{days}$]}&\colhead{}&\colhead{}&\colhead{}&\colhead{}&\colhead{}
} 
\startdata
NGC 4151$^a$ &0.0032 & 5.10 & $1.15\pm0.13$ & 4754 & 8118 &$2.73^{+0.26}_{-0.10}$ & 6.64\% &$2.12$ & 1.62\%&Ref. (1) \\
NGC 7469 & 0.0163 & 1.10 &$ 4.82\pm0.74$ & 5100 & 6868 & $2.40^{+0.22}_{-0.16}$ & 3.66\% &$2.56$ & 5.31\%&Ref. (2)  \\
NGC 3516$^b$  & 0.0088 & 4.73 & 1.2 & 5100 & 4112 & $2.62^{+0.42}_{-0.26}$ & 5.25\% &$2.17$ & 3.58\%&Ref. (3), Ref. (4)   \\
NGC 3516$^c$  & 0.0088 & 4.73 & 0.27 & 4728 & 1085 & $1.75^{+0.16}_{-0.12}$ & 5.15\% &$1.71$ & 4.69\%&Ref. (4), Ref. (5)   \\
\enddata
\tablecomments{Column (1) is object designation; column (2) is the redshift; column (3) is the black-hole mass; column (4) is the bolometric luminosity; column (5) is the rest-frame wavelength; column (6) is the baseline in rest-frame; column (7) is the rest-frame best-fitting damping timescale $\tau_{\mathrm{DRW,obs}}$ from light curves; column (8) is $\tau_{\mathrm{DRW,obs}}$-to-baseline ratio; column (9) is the damping timescale $\tau_{\mathrm{DRW,cal}}$ calculated by Equation \ref{equ: tau_disc-M-mdot}; column (10) is $\tau_{\mathrm{DRW,cal}}$-to-baseline ratio; column (11) is the references: Ref. (1) \citet{Chen2023}, Ref. (2) \citet{Shapovalova2017}, Ref. (3) \citet{Shapovalova2019}, Ref. (4) \citet{Mehdipour2022}, Ref. (5) The g-band data from Zwicky Transient Facility \citep[ZTF;][]{Masci2019}.\\
$^a$ NGC 4151 is a changing-look AGN, and $L_{\mathrm{bol}}$ is taken from the high-flux state. We remove data points with the signal-to-noise ratios less than 20.\\
$^b$ NGC 3516 is also a changing-look AGN, and here are the results in the high-flux state \citep[from 1996 to 2007;][]{Popovic2023}{}{}.\\
$^c$ The results of NGC 3516 in the low-flux state \citep[from 2018 to 2021;][]{Popovic2023}. 
}
\end{deluxetable*}

Having very long light curves is vital to obtain unbiased damping timescales. According to Equation \ref{equ: tau_disc-M-mdot}, to obtain unbiased damping timescales, luminous targets with long-wavelength emission have expected baselines that are decades-long. In contrast, faint counterparts with short-wavelength emission have expected baselines of only a few days to a few years. Therefore, in the case of finite observational baselines, choosing targets with low bolometric luminosity or short rest-frame wavelength emission can improve the reliability of the best-fitting damping timescales. The LSST \citep[][]{Brandt2018}{}{} and WFST \citep[][]{Wang2023-WFST}{}{} will provide a vast amount of AGN optical variability data in the southern and northern celestial hemispheres, respectively. These programs will extend the observational baselines and expand the variability data, helping one to obtain unbiased damping timescales.

Equation \ref{equ: tau_disc-M-mdot} also provides a new method for calculating the absolute accretion rate $\dot{M}$, which $\propto M_{\mathrm{BH}}\dot{m}$. While ensuring that the best-fitting damping timescale is unbiased, $\dot{M}$ should satisfy the following equation, 
\begin{equation}\label{equ: Mdot}
\dot{M}=43.51\tau_{\mathrm{DRW,disk}}^{1.54}\lambda_\mathrm{rest}^{-1.83}\ [\mathrm{M_\odot\ yr^{-1}}].
\end{equation}

\subsection{Variability radius \texorpdfstring{$R_{\mathrm{var}}$}{}}\label{subsec:R_var}
\begin{figure}
    \centering
    \includegraphics{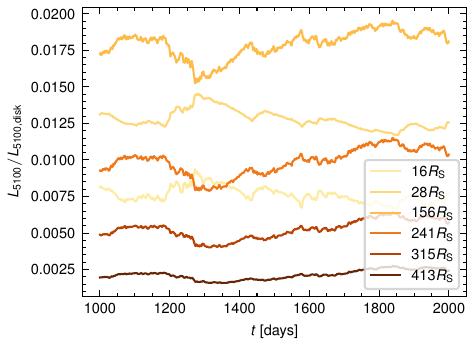}
    \caption{The $5100\ \mathrm{\AA}$ flux variations at different radii of the same accretion disk. The x-axis is time, and the y-axis is the ratio of the luminosity at a given radius to the luminosity of the whole disk at the wavelength of $5100\ \textrm{\AA}$. Darker colors indicate larger radii. Smaller-radii light curves contain more short-term variations than larger-radii ones.}
    \label{fig: lightcurve}
\end{figure}

If the damping timescale is related to the characteristic timescales of the accretion disk, such as the thermal timescale, then according to the static SSD model, the relationship between $\tau_{\mathrm{DRW}}$ and wavelength should be $\tau_{\mathrm{DRW}}\propto\lambda^2$. The observations in S22 concluded $\tau_{\mathrm{DRW}}\propto\lambda^{0.20}$, which is a biased result due to the baseline limitation. In Section \ref{subsec: Mbh&mdot}, we obtain $\tau_{\mathrm{DRW}}\propto\lambda^{1.19}$ after eliminating the effect of baseline inadequacy by simulations, and the dependence of $\tau_{\mathrm{DRW}}$ on wavelength is still weaker than expected from the static SSD model. This is because the best-fitting damping timescale is an average of the radius-dependent characteristic timescales at different radii of the accretion disk. Figure \ref{fig: lightcurve} shows the $5100\ \mathrm{\AA}$ flux variations of a given wavelength at different radii (hereafter, the single-radius light curves) of the same accretion disk. On short timescales, the observed variability is dominated by contributions from inner regions and vice versa. Hence, the best-fitting damping timescale is different from the thermal timescale at which $k_\mathrm{B}T(R_{\lambda})=hc/\lambda$.

\begin{figure*}
    \centering
    \includegraphics[width=27cm,height=10cm,keepaspectratio]{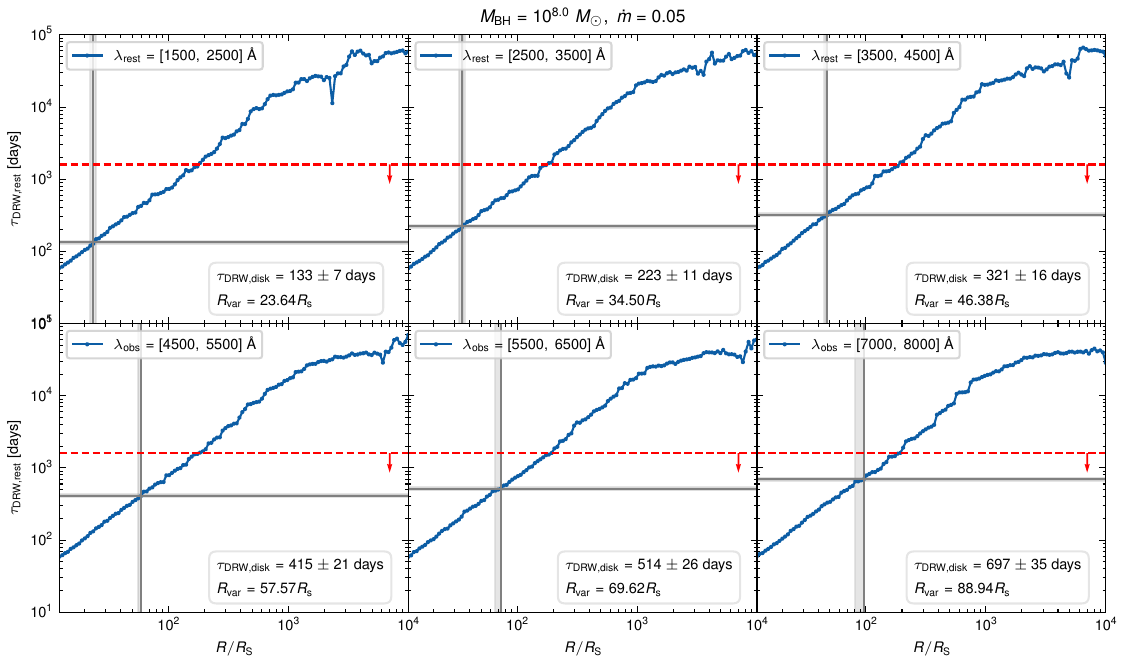}
    \caption{Relation between $\tau_{\mathrm{DRW,radius}}$ and $R/R_\mathrm{S}$ for $M_{\mathrm{BH}}=10^{8.0}\ M_\odot$ and $\dot{m}=0.05$. The six panels represent six bands. Red dashed lines indicate 20\% of the baseline. Horizontal gray lines and shaded gray areas are $\tau_{\mathrm{DRW,disk}}$ and its narrow range, which is 10\% of $\tau_{\mathrm{DRW,disk}}$. Vertical lines and shaded areas indicate a narrow radius range in which $\tau_{\mathrm{DRW,radius}}$ is the same as $\tau_{\mathrm{DRW, disk}}$, and the average of this range is noted as $R_{\mathrm{var}}$.} 
    \label{fig: R_var}
\end{figure*}

We are now interested in finding a characteristic radius (hereafter $R_{\mathrm{var}}$) at which its single-radius light curve has a local damping timescale (i.e., $\tau_{\mathrm{DRW,radius}}$) equaling the damping timescale (i.e., $\tau_{\mathrm{DRW,disk}}$) of the integrated disk light curve. For each case in Table \ref{tab:parameters1} with $L_{\mathrm{bol}}<10^{45.1}\ \mathrm{erg\ s^{-1}}$, in step 4 in Section \ref{subsec:simulation steps}, we generate one hundred sets of single-radius light curves. We fit every single-radius light curve with the DRW model for each band and obtain the corresponding local damping timescale, $\tau_{\mathrm{DRW,radius}}$.
Figure \ref{fig: R_var} represents the relation between $\tau_{\mathrm{DRW,radius}}$ and $R/R_\mathrm{S}$ for $M_{\mathrm{BH}}=10^{8.0}\ M_\odot$ and $\dot{m}=0.05$. 
We define the variability radius $R_{\mathrm{var}}$ to be the characteristic radius at which $\lvert \tau_{\mathrm{DRW, radius}}-\tau_{\mathrm{DRW,disk}} \rvert<0.05\tau_{\mathrm{DRW,disk}}$. 
Then, we find $R_{\mathrm{var}}$ for each case and each band. 

\subsubsection{\texorpdfstring{The dependencies of $\tau_{\mathrm{DRW,disk}}$ upon $R_{\mathrm{var}}/R_{\mathrm{S}}$}{}}\label{subsubsec: taudisc_R-var}
\begin{figure}
    \centering
    \includegraphics{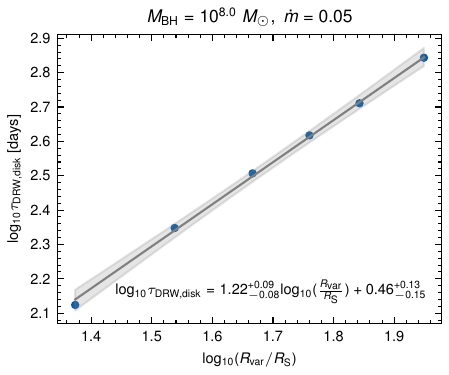}
    \caption{The relationship between $\tau_{\mathrm{DRW,disk}}$ and $R_{\mathrm{var}}$ for $M_{\mathrm{BH}}=10^{8.0}\ M_\odot$ and $\dot{m}=0.05$. The solid line is the best-fitting line and the shaded area is $1\sigma$ confidence intervals.}
    \label{fig: tau_disc-R_var}
\end{figure}

To investigate the relationship between $\tau_{\mathrm{DRW,disk}}$ and $R_{\mathrm{var}}$, we fit $\tau_{\mathrm{DRW,disk}}$ and $R_{\mathrm{var}}$ at different bands for a specific $M_{\mathrm{BH}}$ and $\dot{m}$ using the following equation, 
\begin{equation} \label{equ: tau_disc-R_var}
    \mathrm{log_{10}}\tau_{\mathrm{DRW,disk}}=\alpha\mathrm{log_{10}}(R_{\mathrm{var}}/R_{\mathrm{S}})+\beta.
\end{equation}
For each case in Table \ref{tab:parameters1} with $L_{\mathrm{bol}}<10^{45.1}\ \mathrm{erg\ s^{-1}}$, we repeat the fitting of Equation \ref{equ: tau_disc-R_var} one hundred times and adopt the medians as the parameters $\alpha$ and $\beta$ corresponding to each case. Figure \ref{fig: tau_disc-R_var} represents a fitting result for $M_{\mathrm{BH}}=10^{8.0}\ M_\odot$ and $\dot{m}=0.05$.
The slope differs from the scaling law of $\sim R^{3/2}$. 

\begin{figure*}
    \centering
    \includegraphics[width=21cm,height=6cm,keepaspectratio]{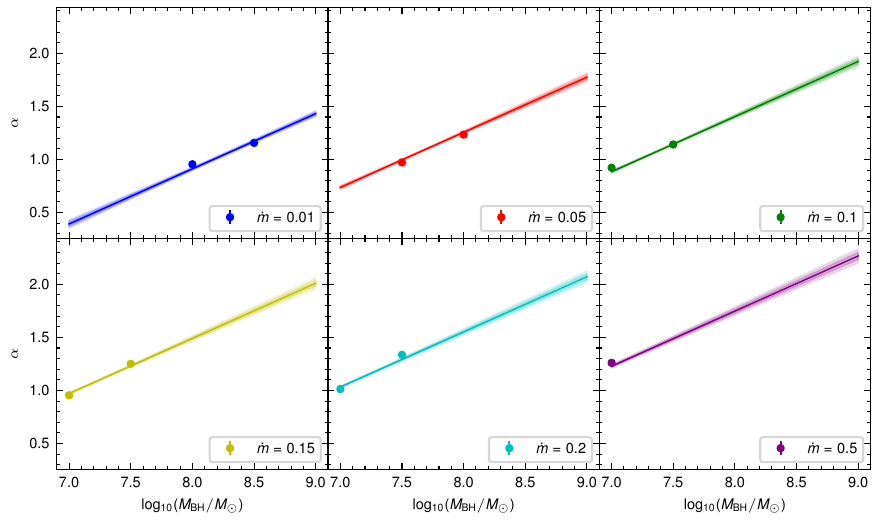}
    \includegraphics[width=21cm,height=6cm,keepaspectratio]{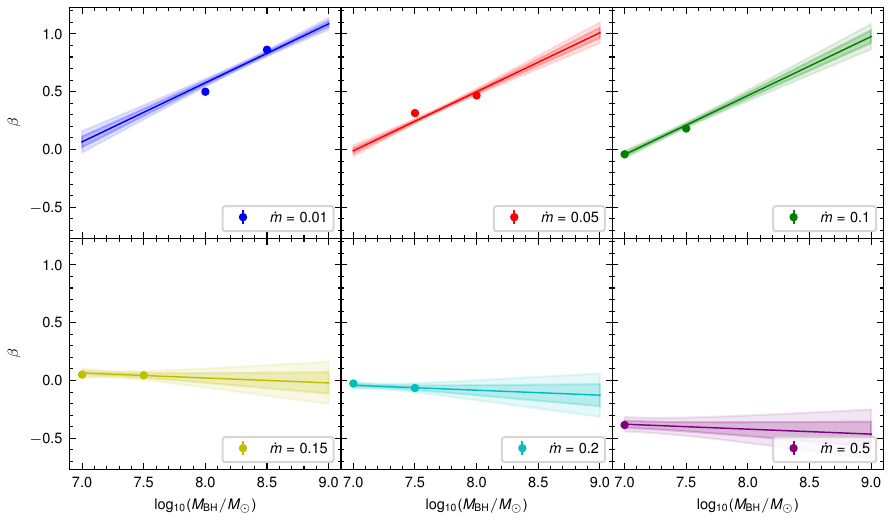}
    \caption{Fitting results (data points) for parameters $\alpha$ and $\beta$ in Equation \ref{equ: tau_disc-R_var}. Different panels represent different $\dot{m}$. We fit all data points with two linear relations for $\alpha$ and $\beta$, respectively. The light- and dark-shaded regions represent the $1\sigma$ and $2\sigma$ confidence intervals of the best-fitting linear relations, respectively.}
    \label{fig: fit alpha and beta}
\end{figure*}

We can obtain $\alpha$ and $\beta$ for each case in Table \ref{tab:parameters1} with $L_{\mathrm{bol}}<10^{45.1}\ \mathrm{erg\ s^{-1}}$ and find that they depend upon $M_{\mathrm{BH}}$ and $\dot{m}$. 
Hence, we aim to find the relationships between the parameters $\alpha$ and $\beta$  and $M_{\mathrm{BH}}$ and $\dot{m}$ (Figure \ref{fig: fit alpha and beta}) by simultaneously fitting these cases. 
The best-fitting results are 
\begin{equation}\label{equ: alpha}
    \alpha = 0.49_{-0.01}^{+0.01}\mathrm{log_{10}}\frac{M_{\mathrm{BH}}}{M_{\odot}}+0.52_{-0.02}^{+0.02}\mathrm{log_{10}}\dot{m}-2.26_{-0.10}^{+0.10},
\end{equation}

\begin{equation}\label{equ:beta}
    \beta = 
    \begin{cases}
        \begin{split}
            0.51_{-0.03}^{+0.03}\mathrm{log_{10}}\frac{M_{\mathrm{BH}}}{M_{\odot}}\!-\!0.11_{-0.04}^{+0.04}\mathrm{log_{10}}\dot{m} & \!-\!3.73_{-0.22}^{+0.22} \\
            &\dot{m}\leq0.1,
        \end{split}
     \\
        \begin{split}
            \!-\!0.04_{-0.05}^{+0.05}\mathrm{log_{10}}\frac{M_{\mathrm{BH}}}{M_{\odot}}\!-\!0.85_{-0.08}^{+0.08}\mathrm{log_{10}}\dot{m} & \!-\!0.33_{-0.36}^{+0.36}  \\
            &\dot{m}>0.1.
        \end{split}
    \end{cases}
\end{equation}
In summary, the relation between the damping timescale and $R_{\mathrm{var}}$ depends upon the black hole mass $M_{\mathrm{BH}}$ and dimensionless accretion ratio $\dot{m}$. 

\subsubsection{\texorpdfstring{The dependencies of $R_{\mathrm{var}}/R_{\mathrm{S}}$ upon $M_{\mathrm{BH}}$, $\dot{m}$, and $\lambda_{{\mathrm{rest}}}$}{}} \label{subsubsec:R_var-Mbh&mdot}
\begin{figure}
    \centering
    \includegraphics[width=18cm,height=6cm,keepaspectratio]{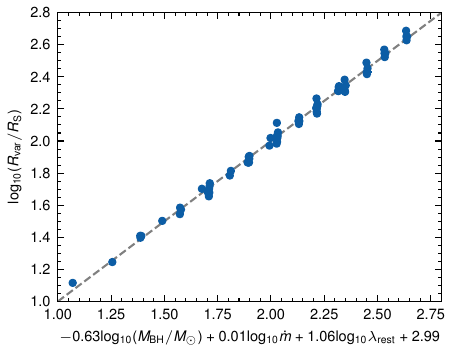}
    \caption{The relationship between $R_{\mathrm{var}}$ and $M_{\mathrm{BH}}$, $\dot{m}$ and $\lambda_{\mathrm{rest}}$. The gray dashed line indicates the one-to-one relation.}
    \label{fig: R_var-M-mdot}
\end{figure}

We also establish the relationship between  $R_{\mathrm{var}}/R_{\mathrm{S}}$ and $M_{\mathrm{BH}}$, $\dot{m}$, and rest-frame wavelength $\lambda_{\mathrm{rest}}$. The fitting equation is
\begin{equation}\label{equ: R_var-M-mdot}
    \begin{split}
        \mathrm{log_{10}}(R_{\mathrm{var}}/R_{\mathrm{S}}) = & u\mathrm{log_{10}}(M_{\mathrm{BH}}/M_{\odot}) +v\mathrm{log_{10}}\dot{m} \\
        & +s\mathrm{log_{10}}(\lambda_{\mathrm{rest}}/\textrm{\AA})+\gamma.
    \end{split}
\end{equation}
Figure \ref{fig: R_var-M-mdot} illustrates the fitting results of Equation \ref{equ: R_var-M-mdot}. The best-fitting parameters are $u\! =\! -\!0.63^{+0.01}_{-0.01}, v\! =\! 0.01^{+0.01}_{-0.01}, s\! =\! 1.06^{+0.01}_{-0.01}, \gamma\! =\! 2.99^{+0.06}_{-0.06}.$ According to Equation \ref{equ: R_var-M-mdot}, $R_{\mathrm{var}}\thicksim\lambda_\mathrm{rest}^{1.06}$, which is less steep than $R_\lambda\thicksim\lambda_\mathrm{rest}^{4/3}$ based on $k_\mathrm{B}T(R_{\lambda})=hc/\lambda$. In previous studies \citep[e.g.,][]{Burke2021}{}{}, it is often argued that the damping timescale should be related to the thermal timescale at $R_{\lambda}$, which scales as $R_{\lambda}^{3/2}\sim\lambda^2$. In the CHAR model, the damping timescale scales as $R_{\mathrm{var}}^{\alpha}\sim\lambda^{s\alpha}$, which is less steep than the scaling relation of $\lambda^2$. 

\subsubsection{\texorpdfstring{The relationship between $R_{\mathrm{var}}/R_{\mathrm{S}}$ and $R_{\mathrm{\lambda}}/R_{\mathrm{S}}$}{}} \label{subsubsec:R_var-R_th}
For a given wavelength $\lambda$, the emission-region size at which $k_\mathrm{B}T(R_{\lambda})=hc/\lambda$ for the CHAR model is \citep[][]{Sun2020}{}{}
\begin{equation}\label{equ: R_th}
    R_{\mathrm{\lambda}}/R_{\mathrm{S}}=(\frac{3(1+k)k_{\mathrm{B}}^4 L_{\mathrm{Edd}}}{64\pi \sigma \eta G^2 M_{\odot}^2 h^4}\lambda^4 (\frac{M_{\mathrm{BH}}}{M_{\odot}})^{-2}\dot{m})^{1/3},
\end{equation}
where $k_\mathrm{B}$, $\sigma$, $G$, $\eta$, and $h$ denote the Boltzmann constant, the Stefan-Boltzmann constant, the gravitational constant, the radiation efficiency (fixed at 0.1), and the Planck constant, respectively.

\begin{figure}
    \centering
    \includegraphics{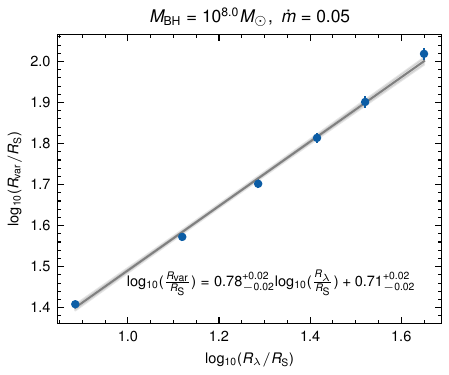}
    \caption{The relationship between $R_{\mathrm{var}}/R_{\mathrm{S}}$ and $R_{\mathrm{\lambda}}/R_{\mathrm{S}}$ for $M_{\mathrm{BH}}=10^{8.0}\ M_\odot$ and $\dot{m}=0.05$. The dots represent our calculations. The solid line and shaded areas are the best-fitting result and $1\sigma$ confidence intervals, respectively.}
    \label{fig: R_var-R_th}
\end{figure}

\begin{figure*}
    \centering
    \includegraphics{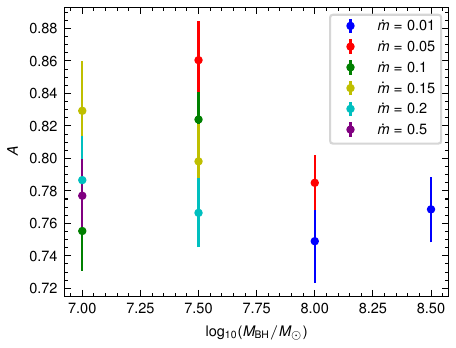}
    \includegraphics{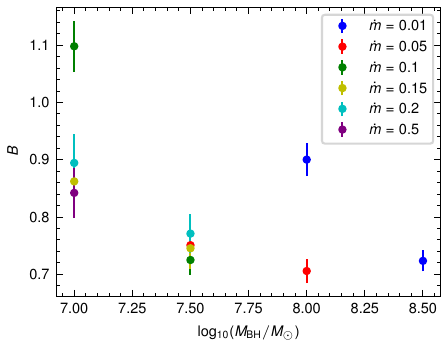}
    \caption{Parameters A and B for different $M_{\mathrm{BH}}$ and $\dot{m}$ in Equation \ref{equ: R_var-R_th}. Different colors represent different $\dot{m}$. 
    \label{fig: A and B}}
\end{figure*}

We try to find the relationship between $R_{\mathrm{var}}$ and $R_{\mathrm{\lambda}}$. We fit $R_{\mathrm{var}}/R_{\mathrm{S}}$ and $R_{\mathrm{\lambda}}/R_{\mathrm{S}}$ using the following relation:
\begin{equation}\label{equ: R_var-R_th}
    \mathrm{log_{10}}(R_{\mathrm{var}}/R_{\mathrm{S}})=A\mathrm{log_{10}}(R_{\mathrm{\lambda}}/R_{\mathrm{S}})+B.
\end{equation}
Figure \ref{fig: R_var-R_th} shows the relationship between $R_{\mathrm{var}}/R_{\mathrm{S}}$ and $R_{\mathrm{\lambda}}/R_{\mathrm{S}}$  for $M_{\mathrm{BH}}=10^{8.0}\ M_\odot$ and $\dot{m}=0.05$. The relationship between $R_{\mathrm{var}}/R_{\mathrm{S}}$ and $R_{\mathrm{\lambda}}/R_{\mathrm{S}}$ is non-linear since $A$ is less than unity. 
Figure \ref{fig: A and B} shows the values of parameters A and B for the different cases. 
It is evident that $A$ and $B$ depend weakly upon $M_{\mathrm{BH}}$ or $\dot{m}$. 
We cannot find a simple function to describe the relationship between $A$ (or $B$) and $M_{\mathrm{BH}}$ or $\dot{m}$. 

\subsubsection{\texorpdfstring{The dependencies of $\tau_{\mathrm{DRW,disk}}$ upon $R_{\mathrm{\lambda}}/R_{\mathrm{S}}$}{}} \label{subsubsec: tau-R_th}
\begin{figure}
    \centering
    \includegraphics{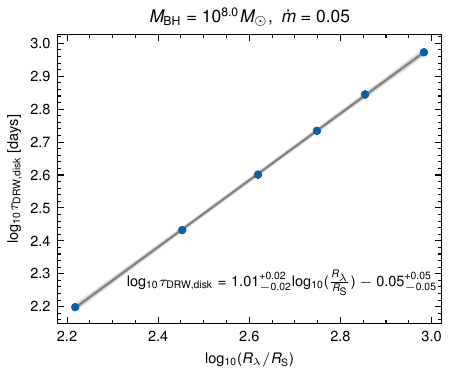}
    \caption{The relationship between $\tau_{\mathrm{DRW,disk}}$ and $R_{\mathrm{\lambda}}/R_{\mathrm{S}}$ for $M_{\mathrm{BH}}=10^{8.0}\ M_\odot$ and $\dot{m}=0.05$. The solid line and shaded areas are the best-fitting relation and $1\sigma$ confidence intervals, respectively.}
    \label{fig: tau_disc-R_th}
\end{figure}

\begin{figure*}
    \centering
    \includegraphics[width=21cm,height=6cm,keepaspectratio]{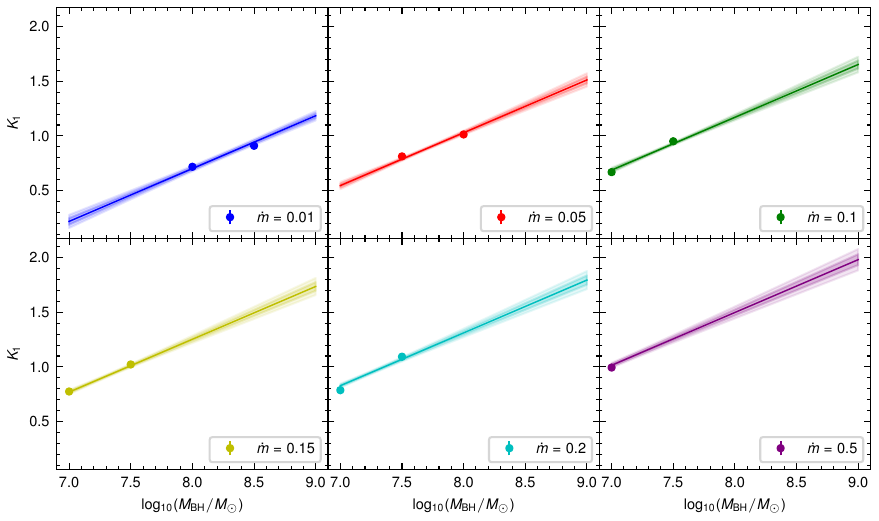}
    \includegraphics[width=21cm,height=6cm,keepaspectratio]{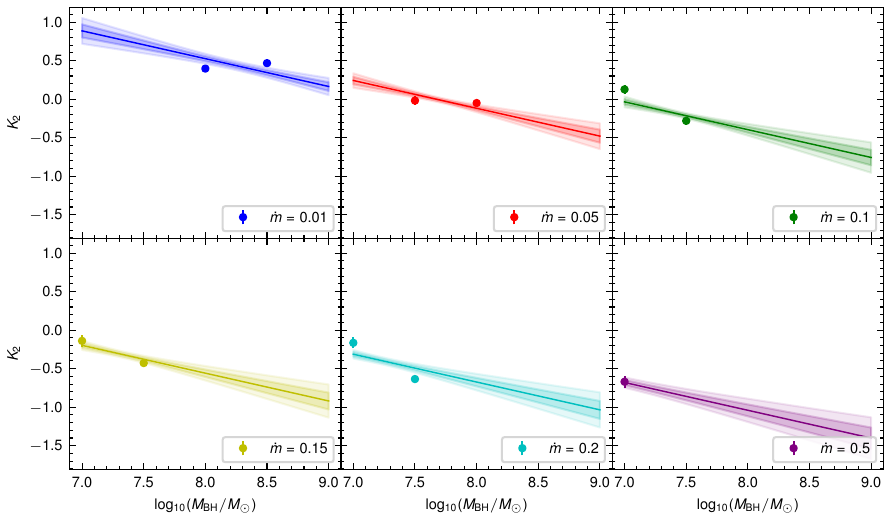}
    \caption{Fitting results (data points) for parameters $K_1$ and $K_2$ in Equation \ref{equ: tau_disc-R_th}. Different panels represent different $\dot{m}$. We fit all data points with two linear relations for $K_1$ and $K_2$, respectively. The light- and dark-shaded regions represent the $1\sigma$ and $2\sigma$ confidence intervals of the best-fitting linear relations, respectively.}
    \label{fig: fit K_1 and K_2}
\end{figure*}

We also aim to establish a relationship between the directly measurable quantity $\tau_{\mathrm{DRW,disk}}$ and $R_{\mathrm{\lambda}}/R_{\mathrm{S}}$. The fitted equation is
\begin{equation}\label{equ: tau_disc-R_th}
    \mathrm{log_{10}}\tau_{\mathrm{DRW,disk}}=K_1\mathrm{log_{10}}(R_{\mathrm{\lambda}}/R_{\mathrm{S}})+K_2.
\end{equation}
Figure \ref{fig: tau_disc-R_th} is the relationship between $\tau_{\mathrm{DRW,disk}}$ and $R_{\mathrm{\lambda}}/R_{\mathrm{S}}$ for $M_{\mathrm{BH}}=10^{8.0}\ M_\odot$ and $\dot{m}=0.05$.
Again, $K_1$ and $K_2$ depend upon $M_{\mathrm{BH}}$ or $\dot{m}$ (Figure~\ref{fig: fit K_1 and K_2}), and the best-fitting relationships are 
\begin{equation}\label{equ: K_1}
    K_1 = 0.47_{-0.02}^{+0.02}\mathrm{log_{10}}\frac{M_{\mathrm{BH}}}{M_{\odot}}+0.48_{-0.02}^{+0.02}\mathrm{log_{10}}\dot{m}-2.22_{-0.16}^{+0.16},
\end{equation}
\begin{equation}\label{equ: K_2}
    K_2 = -0.92_{-0.06}^{+0.06}\mathrm{log_{10}}\frac{M_{\mathrm{BH}}}{M_{\odot}}-0.36_{-0.06}^{+0.06}\mathrm{log_{10}}\dot{m}+1.57_{-0.43}^{+0.43}.
\end{equation}
We stress that $K_1$ is a function of $M_{\mathrm{BH}}$ and $\dot{m}$, rather than being fixed to $3/2$ as expected from the thermal timescale.

\section{Power Spectral Densities at Different Radii} \label{sec: PSD at different radii}
\begin{figure*}
    \centering
    \includegraphics[width=1\textwidth]{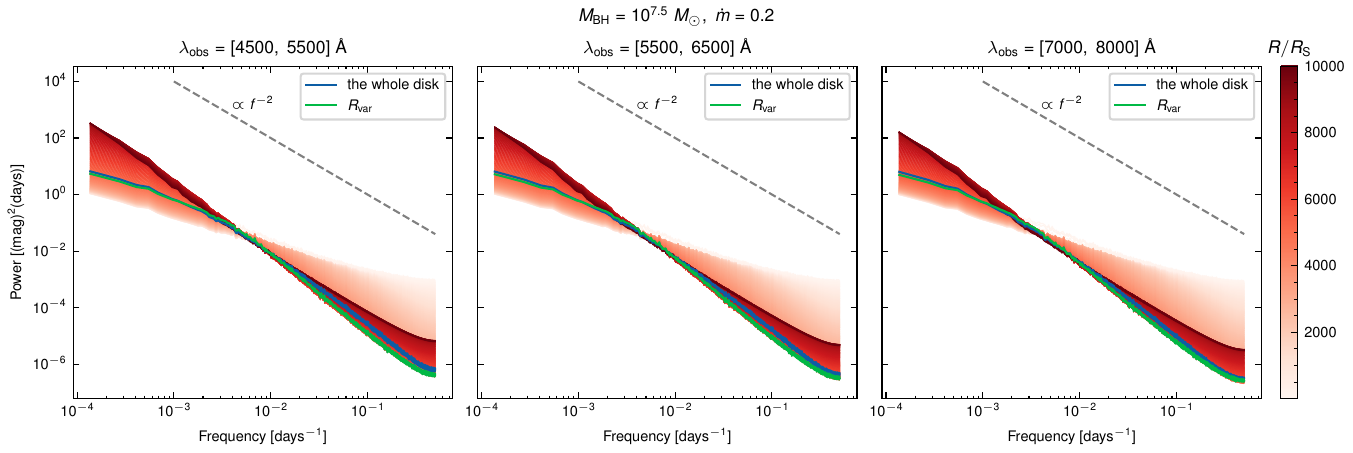}
    \caption{The PSDs at different radii for $M_{\mathrm{BH}}=10^{7.5}\ M_\odot$ and $\dot{m}=0.2$. The red curves represent PSDs at different radii. The blue curves are the PSDs of the whole disk, and the green curves are the PSDs corresponding to $R_\mathrm{var}$.}
    \label{fig: PSD at different radii}
\end{figure*}

We also use the FFT method to calculate the PSDs for the light curves at each radius. Figure \ref{fig: PSD at different radii} shows the PSDs at different radii for a typycal case of $M_{\mathrm{BH}}=10^{7.5}\ M_\odot$ and $\dot{m}=0.2$. It is obvious that the PSD is steeper than that of a DRW on short timescales. This is because, in the CHAR model, the temperature fluctuations have a PSD of $\propto f^{-3}$ on short timescales \citep{Sun2020}. As a result, the PSD of each single-radius light curve is inconsistent with the DRW model. The PSD of the integrated disk light curve is a superposition of various non-DRW PSDs at various disk radii and can resemble the DRW PSD on timescales from months to years \citep[see Figure 4 of][]{Sun2020}. We compare the PSD corresponding to $R_\mathrm{var}$ with that of the whole disk. At high frequencies, the PSDs corresponding to $R_\mathrm{var}$ are smaller than that of the whole disk. The two PSDs are consistent at low frequencies. Hence, the damping timescales from the DRW fitting and PSD analysis should be generally consistent. The PSD at the characteristic radius $R_{\mathrm{var}}$ can roughly represent the PSD of the whole disk. 

\section{Summary} \label{sec: Summary}
We have used the CHAR model to reproduce the observations of S22. In addition, we have obtained a new scaling relation for the intrinsic damping timescale and its connection to the AGN properties through CHAR model simulations. The main conclusions are as follows:
\begin{enumerate}
\item The CHAR model can reproduce the DRW fitting results for the S22 sample, including the best-fitting damping timescales (see Table \ref{tab:tau}), the dependence of the best-fitting damping timescale on wavelength (see Figure \ref{fig: wavelength}, Table \ref{tab:slope}, Section \ref{subsec: wavelength}) and the ensemble PSDs of the sample (see Figures \ref{fig: compare PSD}, \ref{fig: ensPSD by mbh}, and \ref{fig: ensPSD by L}; Section \ref{subsec: PSD}).
\item The observational baselines for most luminous AGNs are not long enough to recover the intrinsic damping timescale. The damping timescale may be biased even if the best-fitting damping timescale is less than 20\% (or 10\%) of the observed baseline (Figure \ref{fig: tau_disc-Lbol}, Section \ref{subsec:simulation steps}). 
\item We have obtained the relationship between the intrinsic damping timescale and $M_\mathrm{BH}$, $\dot{m}$, $\lambda_\mathrm{rest}$ after eliminating the effect of baseline inadequacy by simulations. The main factors affecting the intrinsic damping timescale are the bolometric luminosity and rest-frame wavelength (Equation \ref{equ: tau_disc-M-mdot}, Figure \ref{fig: Stone&Burke}, Section \ref{subsec: Mbh&mdot}). Furthermore, we argue that our results can be used to estimate the absolute accretion rate (see Equation \ref{equ: Mdot}) based on the intrinsic damping timescale. 
\item The observed light curves are a summation of variable emission from various disk radii, and the damping timescale should be some averages of radius-dependent characteristic timescales (see Figure \ref{fig: lightcurve}). This leads to a weaker dependence between the damping timescale and wavelength than the static SSD model. We have obtained the relationship between the directly measurable quantity $\tau_{\mathrm{DRW,disk}}$ and $R_\mathrm{var}$ (see Equation \ref{equ: tau_disc-R_var}, Figures \ref{fig: tau_disc-R_var}, \ref{fig: fit alpha and beta}, Section \ref{subsubsec: taudisc_R-var}), $R_\mathrm{var}$ and $M_\mathrm{BH}$, $\dot{m}$, $\lambda_\mathrm{rest}$ (see Equation \ref{equ: R_var-M-mdot}, Figure \ref{fig: R_var-M-mdot}, Section \ref{subsubsec:R_var-Mbh&mdot}), $R_\mathrm{var}$ and $R_\mathrm{\lambda}$ (see Equation \ref{equ: R_var-R_th}, Figures \ref{fig: R_var-R_th}, \ref{fig: A and B}, Section \ref{subsubsec:R_var-R_th}), and $\tau_{\mathrm{DRW,disk}}$ and $R_\mathrm{\lambda}$ (see Equation \ref{equ: tau_disc-R_th}, Figures \ref{fig: tau_disc-R_th}, \ref{fig: fit K_1 and K_2}, Section \ref{subsubsec: tau-R_th}).
\item  The PSDs corresponding to $R_{\mathrm{var}}$ are consistent with that of the whole disk at low frequencies; the damping timescales from DRW fitting and PSD analysis should be generally consistent (see Figure \ref{fig: PSD at different radii}, Section \ref{sec: PSD at different radii}).
\end{enumerate}
The upcoming LSST and WFST are expected to provide tremendous AGN variability data that will enlarge the light-curve baselines significantly, further validating our results.

\begin{acknowledgments}
We thank Pu Du for offering the light curve of NGC 4151 and for beneficial discussions. We thank the referee for his/her helpful suggestions that improved the manuscript. S.Y.Z. and M.Y.S. acknowledge support from the National Natural Science Foundation of China (NSFC-12322303; NSFC-11973002), the Natural Science Foundation of Fujian Province of China (No. 2022J06002), and the China Manned Space Project grant (No. CMS-CSST-2021-A06). Z.Y.C. and J.X.W. acknowledge support from the National Natural Science Foundation of China (NSFC-12033006). Y.Q.X. acknowledges support from the National Natural Science Foundation of China (NSFC-12025303; NSFC-11890693). 

Based on observations obtained with the Samuel Oschin Telescope 48-inch and the 60-inch Telescope at the Palomar Observatory as part of the Zwicky Transient Facility project. ZTF is supported by the National Science Foundation under Grants No. AST-1440341 and AST-2034437 and a collaboration including current partners Caltech, IPAC, the Weizmann Institute for Science, the Oskar Klein Center at Stockholm University, the University of Maryland, Deutsches Elektronen-Synchrotron and Humboldt University, the TANGO Consortium of Taiwan, the University of Wisconsin at Milwaukee, Trinity College Dublin, Lawrence Livermore National Laboratories, IN2P3, University of Warwick, Ruhr University Bochum, Northwestern University and former partners the University of Washington, Los Alamos National Laboratories, and Lawrence Berkeley National Laboratories. Operations are conducted by COO, IPAC, and UW.
\end{acknowledgments}

\vspace{5mm}

\facilities{ZTF \citep{ZTF-DOI}}

\software{Astropy \citep{astropy}, emcee \citep{emcee}, Matplotlib \citep{matplotlib}, Numpy \citep{2020NumPy-Array}, Scipy \citep{2020SciPy-NMeth}, taufit \citep{Burke2021}.}

\bibliography{ref.bib}{}
\bibliographystyle{aasjournal}

\end{document}